\documentclass[11pt,cits]{JHEP3}

\usepackage{epsfig}
\usepackage{latexsym}
\usepackage{graphicx}
\usepackage{amsmath,amssymb}
\usepackage{epstopdf}  
\def\unit{\relax{\rm 1\kern-.26em I}}

\def\bea{\begin{eqnarray}}
\def\eea{\end{eqnarray}}
\def\be{\begin{equation}}
\def\ee{\end{equation}}
\def\nn{\nonumber}

\newcommand{\gsim}{\lower.7ex\hbox{$\;\stackrel{\textstyle>}{\sim}\;$}}
\newcommand{\lsim}{\lower.7ex\hbox{$\;\stackrel{\textstyle<}{\sim}\;$}}

\title{Moduli Stabilization in non-Supersymmetric Minkowski Vacua with Anomalous $U(1)$ Symmetry}
\author{Diego Gallego and Marco Serone \\
International School for Advanced Studies (SISSA/ISAS)
and INFN, Trieste, Italy\\
E-Mail: \email{gallego@sissa.it,serone@sissa.it}}

\abstract{We study  how two moduli can be stabilized in a Minkowski/de Sitter vacuum for a 
wide class of string-inspired Supergravity models with an effective Fayet-like Supersymmetry breaking. 
It is shown under which conditions this mechanism can be made natural and how it can give rise
to an interesting spectrum of soft masses, with a relatively small mass difference between scalar and gaugino masses.
In absence of a constant superpotential term, the above mechanism becomes completely natural and gives rise
to a dynamical supersymmetry breaking mechanism. Some specific type IIB and heterotic string inspired models are considered in detail.
}

\preprint{SISSA-40/2008/EP}


\begin{document}

\section{Introduction}

Stabilizing all moduli in a Minkowski vacuum with low energy supersymmetry (SUSY) breaking
is among the most important problems in string theory, being a crucial step to connect string theory
with SUSY phenomenology, that is supposed to be the best motivated scenario for new physics beyond the Standard Model.
In recent years there has been a tremendous progress on the issue of stabilizing moduli, mostly in Type II string theories and at the supergravity (SUGRA) level. 
A combination of fluxes for Ramond--Ramond (RR) tensor field strengths and for the Neveu Schwarz--Neveu Schwarz (NSNS)  field strength $H$ has been shown  to stabilize the complex structure moduli of the unperturbed compactified space,  typically a Calabi-Yau 3-fold for ${\cal N}=1$ SUSY theories in $D=4$ dimensions  \cite{Giddings:2001yu}. Subsequently  a ``scenario'' with the K\"ahler structure moduli stabilized on a Minkowski/de Sitter (dS) vacuum with SUSY breaking has been introduced by Kachru, Kallosh, Linde and Trivedi (KKLT) \cite{KKLT}, assuming a complete decoupling between complex structure and K\"ahler structure moduli. In addition to this assumption, 
 the authors of \cite{KKLT} invoked an explicit SUSY breaking mechanism (the introduction of ${\bar D}_3$ branes).
Although such explicit breaking does not forbid a quantitative study of some interesting physical quantities, such as soft parameters 
(see e.g.~\cite{Choi:2004sx}), it is nevertheless desirable to have a fully satisfactory spontaneous SUSY breaking mechanism,   motivated also by the fact that SUSY is a local symmetry at the SUGRA level and hence
explicit breakings should be avoided. After the work of \cite{KKLT}, indeed, many works have appeared where the explicit SUSY breaking by ${\bar D}_3$ branes is replaced by spontaneous $F$ or $D$-term breaking of different kind \cite{Endo:2005uy,Villadoro:2005yq,Achucarro:2006zf,GomezReino:2006dk,Parameswaran:2006jh,Lebedev:2006qq,Choi:2006bh,Brummer:2006dg,Dudas:2006vc,Dudas:2006gr,AHKO,OKKLT2,Lebedev:2006qc,Cremades:2007ig,Lalak:2007qd,Dudas:2007nz,Achucarro:2007qa}.

A well-known, simple and interesting SUSY breaking mechanism is the one by Fayet and  
based on a Fayet-Iliopoulos (FI) term for a $U(1)_X$ gauge symmetry \cite{Fayet:1974jb}. Its simplest implementation 
requires two charged fields, $\phi$ and $\chi$, with opposite $U(1)_X$ charges $q_\phi$ and $q_\chi$. The requirement
of minimizing the $D_X$ term in the scalar potential induces one of the charged fields, say $\phi$, to get
a non-vanishing vacuum expectation value (VEV).  If  the only relevant superpotential term coupling $\phi$ and $\chi$ is 
linear in $\chi$, a simple effective Polonyi-like superpotential term is induced and SUSY is broken because  $F_\chi\neq 0$.
A constant FI term necessarily requires in SUGRA the introduction of non-gauge invariant superpotentials, which 
do not seem to occur in string theory.  On the contrary, field-dependent effective FI terms generally arise
due to a non-linear transformation of some modulus $U$ under a would-be anomalous $U(1)_X$ gauge symmetry. 
Moduli stabilization and Fayet-like SUSY breaking mechanisms are hence closely interconnected in string theory and
one might wonder if their combined action can efficiently be embedded in a KKLT-set up to provide a spontaneous
SUSY breaking mechanism, which also  does not need to assume a complete decoupling between moduli stabilization
and SUSY breaking, like in \cite{KKLT}.
A Fayet-like SUSY breaking mechanism has been already shown to successfully give rise to low energy SUSY breaking on a Minkowski/dS vacuum for a KKLT-like SUGRA model, where the FI  modulus is identified with the universal K\"ahler modulus and $q_\phi = -q_\chi$ \cite{Dudas:2007nz}.
The resulting soft mass parameters for the visible sector are realistic, but present a moderate hierarchy between gaugino and scalar masses, unless one complicates the model by introducing additional (messenger-like) fields \cite{Dudas:2007nz}. 
The main drawback of the model of \cite{Dudas:2007nz} is the introduction of an unnaturally small mass term $m\phi \chi$ with  $m\sim O(10^{-11}\div 10^{-12})$,  in addition to the usual KKLT fine-tuning of assuming a tiny constant superpotential term $w_0$, which is roughly of the same order as $m$. A more satisfactory explanation of the smallness of the $\chi \phi$ coupling is necessary, mainly because higher-order terms of the form $c_n (\phi \chi)^n$ with $n>1$ will generally lead to a restoration of SUSY.

 Aim of this paper is to study in some detail how a Fayet-like SUSY breaking mechanism can be realized in string-derived SUGRA theories.
We consider a bottom-up four dimensional ${\cal N}=1$ SUGRA framework, more general than the Type IIB KKLT-like SUGRA setting, so that
our results will be of more general validity. In particular we will study the dynamics of a SUGRA system with two moduli, the FI modulus $U$ and another neutral modulus $Z$, 
the two charged fields $\phi$ and $\chi$, with arbitrary $q_\chi$ charge (the $U(1)_X$ charges are normalized so that $q_\phi = -1$), extra hidden vector-like matter and two/three condensing gauge groups responsible for non-perturbatively generated terms necessary to stabilize $U$ and $Z$.
We specify only the schematic form of the K\"ahler potential,  which is taken quite generic. As far as the superpotential is concerned, we consider both the cases of moduli-independent couplings of the form $Y \chi \phi^{q_\chi}$ and
non-perturbatively generated moduli dependent couplings of the form $Y \exp(-\gamma_U U -\gamma_Z Z) \chi \phi^{q_\chi}$,
where $\gamma_{U,Z}$ are some unspecified constants. 
Once the hidden mesons of the condensing gauge groups are integrated out,
the superpotential becomes the sum of exponential terms (including a constant term $w_0$) and of the coupling
$\chi \phi^{q_\chi}$.  In order to have as much as possible analytic control on this complicated system,  
we first look for SUSY vacua when $\chi$ is decoupled and then we consider its backreaction, which generally gives rise
to non-SUSY vacua. In this approximation, the coupling  $\chi \phi^{q_\chi}$ is effectively an ``up-lifting'' term, required
to pass from the AdS SUSY vacuum to a Minkowski/dS one with low-energy SUSY breaking. 
We find that the last step puts quite severe constraints on the parameter space of the superpotential.
In particular, we find that the non-perturbatively generated couplings can alleviate the tuning needed on $Y$ from 2 to 6 orders of magnitude, depending on the model considered, but taken alone they do not allow us to have $Y\sim O(1)$, since
the back-reaction induced by the up-lifting term becomes so large that the non-SUSY vacuum either disappears or always remains  AdS.
Natural values of $Y$ can however be obtained by assuming that $|q_\chi|>q_\phi$ by a few, so that $\langle \phi \rangle^{q_\chi}$
can be responsible for the remaining necessary suppression. 

Most of our interest is in the hidden sector dynamics of the theory and hence we will not systematically study how SUSY 
breaking is mediated to the visible sector or the precise form of the soft parameters, most of which are necessarily model dependent.
We just notice that gravity mediation of SUSY breaking is preferred to avoid small moduli masses, linked to the gravitino mass, and that the general pattern of the soft mass terms seem very promising. In particular, the gaugino masses, which 
in string-inspired models with purely gravity mediated SUSY breaking are often considerably smaller than the non-holomorphic scalar masses, can be naturally made heavier in our set-up, thanks to the presence of  two moduli.
In presence of the FI modulus only, gauginos can take a mass only by assuming a $U$-dependent gauge kinetic function for the visible gauge group.
Since $U$ transforms under a $U(1)_X$ gauge transformation, anomaly cancellation arguments require that some visible
matter field has to be charged under $U(1)_X$, which in turn gives rise to heavy non-holomorphic scalar masses, induced by the $D_X$ term \cite{Dudas:2007nz}. This problem is now simply solved by assuming that the gauge kinetic function for the visible sector
depends on $Z$ only. Moreover, when $\gamma_Z\neq 0$, the back-reaction of the up-lifting term on $Z$ will give rise
to an enhancement of $F_Z$, so that eventually gauginos just a few times lighter than the gravitino can 
be obtained.

The fine-tuning problem to get $Y\ll 1$, by means of the flatness condition,  is just a reflection of the 
other fine-tuning problem which requires $w_0\ll 1$ to get low-energy supersymmetry breaking. 
Motivated by the idea of solving this second tuning problem, we have also analyzed the situation in which $w_0$ vanishes, 
assuming that some stringy symmetry forbids its appearance. 
In this case, the moduli stabilization mechanism boils down to a racetrack model \cite{Krasnikov:1987jj}, 
where the scale of supersymmetry breaking is dynamically generated.  
Interestingly enough, the back-reaction of the up-lifting term is milder than before and
it is now possible to achieve $Y\sim O(1)$ with small $q_\chi$ charges,  relying on $\gamma_{Z,U}$ only.
Aside from the usual cancellation of the cosmological constant, this model is hence completely natural.
When two moduli are considered, similarly to before, one can have a not too hierarchical spectrum of soft masses,
although the gauginos are now a bit lighter than in the models with $w_0\neq 0$ considered before.

We have supplemented all the above general considerations by analyzing in some detail three specific models:
i) an orientifold compactification of type IIB on $\textbf{CP}^4_{[1,1,1,6,9 ]}$, where $U$ and $Z$ are identified with the 
two K\"ahler moduli of the compactification manifold ($w_0\neq 0$), ii) an heterotic  compactification on a generic Calabi-Yau 3-fold, where
$U$ and $Z$ are identified with the dilaton and the universal K\"ahler modulus, respectively ($w_0\neq 0$), iii) the model of i) but now assuming $w_0=0$.
It should be stressed that, in the spirit of our bottom-up approach,
none of these models are actually full-fledged string models, but rather they should be seen as inspiring examples to partially
fix the arbitrariness in the choice of the K\"ahler potential, superpotential and gauge kinetic functions we have at the SUGRA level.
In all three cases, we have analytically and numerically studied the main properties of the vacuum, including its (meta)stability.
The latter is essentially never a serious issue and all the moduli components
are always one or two order of magnitudes heavier than the gravitino. In all three cases we have always estimated under which conditions
a classical SUGRA analysis is reliable by considering the effect of the universal $\alpha^\prime$ correction to the moduli
K\"ahler potential \cite{Candelas:1990rm,Becker:2002nn}.

The paper is organized as follows. In section 2 we introduce our general model and analytically show under which conditions
a non-SUSY Minkowski/dS vacuum can be obtained. We also show in subsection 2.2
how an effective description where the $U(1)_X$ gauge field
and the field $\phi$ are integrated out is very useful to get some understanding about some properties of the theory. In subsection 3 we analyze the models with $w_0=0$, where SUSY breaking is dynamically generated and the up-lifting term is completely natural. In section 4 a brief discussion 
on the soft mass terms is reported. Section 5 is devoted to the analysis of the example of models i), ii) and iii) mentioned earlier. Section 6 contains our conclusions. Throughout all the paper we use units in which the reduced Planck mass $M_{P}=1$.

\section{General Two-Moduli Model}\label{UV-Heterotic-sect}

The SUGRA model we consider consists of two moduli multiplets
$U$ and $Z$, a hidden gauge group of the form
$G_1\times G_2\times U(1)_X$, with $G_1$ and $G_2$  non-abelian
factors, massless matter charged  under $U(1)_X$ and under $G_1$
or $G_2$ (but not both) and finally two chiral multiplets $\phi$
and $\chi$, charged under $U(1)_X$ and singlets under $G_1\times
G_2$ . For concreteness, we take $G_i=SU(N_i)$ ($i=1,2$) and
consider $N_{fi}$ quarks $Q_i$ and $\widetilde Q_i$ in the
fundamental and anti-fundamental representations of $G_i$. This is
the field content of our model. We assume that the $U(1)_X$ gauge
symmetry is ``pseudo''-anomalous, namely that such symmetry
is non-linearly realized in one of the two moduli multiplets, $U$,
the latter mediating a generalized Green-Schwarz mechanism \cite{Dine:1987xk}.
We normalize the $U(1)_X$ charges so that $q_\phi = -1$ and take $q_{Q_i}+q_{\widetilde
Q_i} >0$, $q_\chi>0$, with the same $U(1)_X$ charge for all
flavours, for simplicity. The model is finally specified by the
K\"ahler potential, superpotential and gauge kinetic functions.
Omitting for simplicity flavour and color indices, the full K\"ahler
potential $K_{Tot}$ and superpotential $W_{Tot}$ are a sum of a visible
and hidden sector,  $W_{Tot}=W_v + W_h$ and $K_{Tot}=K_v+ K_h$,
where
\be
K_v  =   \alpha_{iv} Q_{iv}^{\dagger} e^{2q_{iv} V_X + V_v} Q_{iv} \label{Kahlervisible} 
\ee
represents the K\"ahler potential of the visible sector, with $i$ running over all visible fields,  and we have schematically denoted by $Q_{iv}$ and $V_v$ all the visible chiral fields and vector superfields. 
For simplicity, we have taken $K_v$ to be diagonal in the visible sector fields.
We do not specify the visible superpotential $W_v$ because it will never enter
in our considerations. 
The hidden sector K\"ahler and superpotential terms 
read\footnote{See also \cite{Choi:2006bh} where a similar analysis with a single modulus in a KKLT context has been done.}
\bea
\hspace{-2cm} K_h & = &  K_M+\alpha_\phi \phi^\dagger e^{- 2V_X} \phi+ \alpha_\chi \chi^\dagger e^{ 2 q_\chi V_X} \chi  \nn \\
&&\hspace{1cm} +  \sum_{i=1,2}\alpha_i \Big( Q_i^\dagger e^{V_i +2 q_{Q_i} V_X} Q_i  + \widetilde Q_i^\dagger e^{-V_i +2 q_{\widetilde Q_i} V_X}\widetilde  Q_i \Big) \,,  \label{fullKuv} \\
 W_h &  = &  w_0 + Y(U,Z,\phi) \phi^{q_\chi} \chi  + \sum_{i=1,2} c_i(U,Z,\phi) \,  Q_i \widetilde Q_i \,\phi^{ q_{Q_i}+q_{\tilde Q_i}} \nn \\
&&\hspace{0.2cm}  +\sum_{i=1,2} \eta_i (N_i - N_{if}) \bigg(\frac{\Lambda_i(U,Z)^{3N_i-N_{if}}}{\det(Q_i \widetilde Q_i) }\bigg)^{\frac{1}{N_i-N_{if}}}\,.
 \label{fullWuv}
\eea
The holomorphic gauge kinetic functions in the hidden sector are taken to be
\be
 f_i(U,Z)  =  n_i U + m_i Z + p_i \,, \ \ \ \ \ \ \ f_X(U) = n_X U\,. \label{gaugekinetic}
\ee
Several comments are in order. The K\"ahler potentials (\ref{Kahlervisible}) and
(\ref{fullKuv}) are supposed to be the first terms in an expansion in the matter fields up to quadratic order;
 $K_M,\alpha_{iv}$,  $\alpha_i$, $\alpha_\phi$ and $\alpha_\chi$\footnote{Notice that for simplicity we have taken the same moduli dependent functions $\alpha_i$ for the hidden quarks and anti-quarks.} are
generally real functions of $U+U^\dagger-\delta V_X$ and $Z+Z^\dagger$. In $K_h$,  $V_i$ and $V_X$ denote the vector superfields
associated to the non-abelian groups $G_i$ and $U(1)_X$,
respectively,
$K_M$ is the K\"ahler potential for the $U$ and $Z$ moduli
and $\delta$ is the Green-Schwarz coefficient. The form of the latter is uniquely fixed by
gauge invariance to be
\be
\delta = \frac{(q_{Q_1}+q_{\widetilde Q_1}) N_{1f}}{4\pi^2 n_1} = \frac{(q_{Q_2}+q_{\widetilde Q_2})
N_{2f}}{4\pi^2 n_2}\,. 
\label{GSexplicit}
\ee
In the superpotential (\ref{fullWuv}), we have allowed for an arbitrary constant term $w_0$ that is supposed to be the
left-over of all the remaining fields which are typically present in any explicit string model --- such as the complex structure moduli in KKLT-like compactifications --- and 
assumed of having been integrated out.\footnote{Strictly speaking,  these fields can also generate a constant K\"ahler potential term. For simplicity we neglect it, since it just corresponds to a rescaling of the gravitino mass.\label{footk0}}  We assume that $w_0$ is tiny in Planck units, in order to give rise to a light enough gravitino mass.\footnote{The gravitino mass might also be suppressed by a large negative
K\"ahler potential term, like in the large-volume models of \cite{Balasubramanian:2005zx}, so that $w_0\sim O(1)$ can be taken. We do not consider this possibility here.}
We will later discuss  the case in which $w_0$ exactly vanishes. The hidden Yukawa couplings $Y$ and $c_i$ in $W_h$ are assumed to generally depend on both moduli. Due to the Peccei-Quinn symmetries associated to  ${\rm Im}\, U$ and ${\rm Im}\,Z$, the only allowed moduli dependence is exponential.  The superpotential (\ref{fullWuv}) is
manifestly $G_i$ invariant, whereas the $U(1)_X$ invariant is less
transparent. Under a $U(1)_X$ super-gauge transformation with
parameter $\Lambda$, one has $\delta_X V_X =  -i
(\Lambda-\bar\Lambda)/2$, $\delta_X U = i \delta \Lambda/2$ and
$\delta_X \Phi = i  q_\Phi \Lambda \Phi$ for any charged multiplet
$\Phi$. Gauge invariance constraints then the couplings $Y$ and $Z$ to depend on
$U$ by means of the gauge invariant combination $\exp(-U)\phi^{\delta/2}$.
We then parameterize
\be Y(U,Z,\phi) = Y \phi^{\gamma_U\delta/2} e^{-\gamma_U U - \gamma_Z Z} \,, \ \ \ \
\ \ c_i(U,Z,\phi) = c_i \phi^{\eta_{i,U} \delta/2} e^{-\eta_{i,U} U-\eta_{i, Z} Z}\,, \label{yukawauv}
\ee
with $Y$ a constant and $c_i$ constant matrices (in flavour
space).  The phenomenological coefficients $\gamma_{U/Z}$ and $\eta_{i,U/Z}$ are non-vanishing
for non-perturbatively generated Yukawa couplings only. 
The last term in $W_h$ is the non-perturbatively generated superpotential term
appearing in ${\cal N}=1$ theories for $N_f<N$  \cite{Affleck:1983mk}.
 We have found convenient to introduce the factors $\eta_{1,2}=\pm 1$
in eq.(\ref{fullWuv}),  which will allow us to set to zero the imaginary parts of $U$ and $Z$.
The dynamically generated scales $\Lambda_i(U,Z)$ are field-dependent and follows from the holomorphic gauge
kinetic functions (\ref{gaugekinetic}). From eq.(\ref{gaugekinetic}) we have
\be g_i^{-2} =  {\rm Re} \,f_i(U,Z), \ \ \  g_X^{-2} =  {\rm Re} \, f_X(U)\,,
\ee
and
\be |\Lambda_i(U,Z)| = e^{-\frac{8 \pi^2}{g_i^2
(3N_i-N_{if})}} \,. \label{DynGenScale}
\ee
The coefficients $n_X$, $n_i$, $m_i$ and $p_i$  in eq.(\ref{gaugekinetic}) are model dependent constants, which we keep generic for the moment.  Just for simplicity of the analysis, we have assumed that the $U(1)_X$ factor depends only
on the $U$ modulus. It is straightforward to check that the non-perturbative superpotential terms in
eq.(\ref{fullWuv}) are $U(1)_X$ gauge-invariant provided the
two equalities in eq.(\ref{GSexplicit}) are satisfied.

We will mostly be interested in the dynamics of the hidden sector of the theory, assuming that
all visible fields vanish.
The scalar potential of the theory has the usual SUGRA form given by $V=V_F+V_D$, with
\bea
V_F & = &  e^{K_h} \Big(K_h^{I\bar J}D_I W_h \overline{D_{J} W_h}  -3|W_h|^2\Big) \, , \label{fullVF} \\
V_D & = &  \sum_{i=1,2}\frac{1}{2{\rm Re}\, f_i} D_i^2 +
\frac{1}{2{\rm Re}\, f_X} D_X^2 \,. \label{fullVD} \eea In
eq.(\ref{fullVF}), $I,J$ run over the hidden chiral multiplets $Q_i, \widetilde Q_i, \phi, \chi,U,Z$, $D_I W_h  = \partial_I W_h + (\partial_I K_h) W_h\equiv F_I$ is the
K\"ahler covariant derivative and $K_h^{I\bar J}$ is the inverse
K\"ahler metric. In eq.(\ref{fullVD}), $D_i$ and $D_X$ are the
D-terms associated to the $G_i$ and $U(1)_X$ isometries  of $K$,
generated by Killing vectors $X_i$ and $X_X$, and are given by
(omitting gauge indices) \be D_i  =   \frac{X^I_i F_I}{W} = X^I_i
\partial_I K \,, \ \ \ \ D_X  =    \frac{X^I_X F_I}{W} = X^I_X
\partial_I K \,,\label{FDrel} \ee where the second equalities in the
two expressions above apply for a  gauge invariant
super-potential. The explicit form of the D-terms
is\footnote{Below and throughout the paper, we use the same
notation to denote a chiral superfield and its lowest scalar
component, since it should be clear from the context to what we
are referring to.}
\bea
D_i^a & = & \alpha_i  (Q_i^\dagger T^{a_i} Q_i -  \widetilde Q_i^\dagger T^{a_i} \widetilde Q_i )
\,, \label{Diterm} \\
D_X & = &   \sum_{i=1,2}\alpha_i \Big(q_{Q_i} Q_i^\dagger Q_i
+q_{\widetilde Q_i} \widetilde Q_i^\dagger \widetilde Q_i \Big) +
\alpha_\chi q_\chi \chi^\dagger \chi -\alpha_\phi \phi^\dagger \phi \nn \\
&& -\frac{\delta}{2} \Big[ \alpha_i^\prime  (Q_i^\dagger Q_i
+ \widetilde Q_i^\dagger \widetilde Q_i ) +\alpha_\phi^\prime  \phi^\dagger \phi
+ \alpha_\chi^\prime  \chi^\dagger \chi +  K_M^\prime\Big] \,, \label{Dxterm}
\eea
where $T^{a_i}$ in eq.(\ref{Diterm}) are the generators of $SU(N_i)$ and $\prime$ in eq.(\ref{Dxterm}) stands for a derivative with respect to $U$.

\subsection{Looking for non-SUSY Minkowski minima}

A direct analytical study of the minima of $V$ is a formidable task.
However, we will see that it is possible to find non supersymmetric
metastable Minkowski minima starting from AdS SUSY vacua when $\chi\ll 1$.

Particularly important for what follows is the $U(1)_X$ D-term. As well-known, the GS coefficient $\delta$ induces
a (field-dependent) FI term in $D_X$, as can be seen from eq.(\ref{Dxterm}). 
The minimization of $D_X$ induces then a non-vanishing VEV for $\phi$ (taken real for
simplicity):
\be
\phi_{SUSY}^2 =  \frac{-\delta K_M^\prime}{2( \alpha_\phi+\delta \alpha_\phi^\prime/2 )}\,.
\label{phi0}
\ee
Notice that typically $K_M^\prime <0$, so that the right-hand side in eq.(\ref{phi0}) is positive and the $U(1)_X$ symmetry is spontaneously broken.
From the third term in the superpotential (\ref{fullWuv}), we see that $\phi_{SUSY}$  also induces a
mass term for the quarks $Q_i$ and $\widetilde Q_i$.
Assuming that $ \phi_{SUSY}\gg m_{3/2}$,  unless the Yukawa couplings $c_i(U,Z,\phi)$ are extremely small,
a sufficiently large mass  for the quarks  $Q_i$ and $\widetilde
Q_i$ is induced.\footnote{We assume that the $c_i$ in eq.(\ref{yukawauv}) are such that all quarks get a mass when $\phi$ acquires a VEV.} Under the assumption 
that at the minimum $W_h\ll 1$, which is obviously required to have a sufficiently light gravitino mass, the quarks can be integrated out by safely neglecting
all supergravity and moduli corrections, by setting to zero $D_i$ and their flat-space F-terms: $F_{M_i} =
\partial W_h/\partial M_i = 0$,  where $M_i = Q_i \widetilde Q_i$ are the ``meson''  chiral fields.
The effective superpotential $W_{eff}$ which result after having integrated out the quark superfields reads
\be
W_{eff} = w_0 + f(\phi) e^{- \gamma_Z Z-\gamma_U U } \chi + \sum_{i=1,2} A_i(\phi) e^{-a_i U - b_i Z}\,, \label{Weff}
\ee
where
\bea
a_i  \equiv   \eta_{i,U} \frac{N_{if}}{N_i}+ \frac{8\pi^2 n_i}{N_i} \,, \ \ \ \   b_i   \equiv    \eta_{i,Z} \frac{N_{if}}{N_i}+ \frac{8\pi^2 m_i}{N_i}\,,\nn \\
f(\phi)   \equiv   Y \phi^{\hat q_\chi}\,,  \ \ \ \ A_i (\phi) \equiv  \eta_i N_i e^{-8\pi^2 p_i/N_i} \Big( c_i\phi^{ q_i} \Big)^{\frac{N_{if}}{N_i}} \!,
\label{abcoeff}
\eea
are effective parameters and for simplicity we have defined the effective charges
\be
q_i\equiv q_{Q_i}+q_{\tilde Q_i}+\eta_{i,U}\delta/2\,, \ \ \ \ \hat q_\chi \equiv q_\chi + \frac{\gamma_U \delta}{2}\,.
\ee
The meson VEV's are negligibly small, in agreement with our assumption:
\be
\langle M_i \rangle = \Lambda_i^2 \bigg( \frac{\Lambda_i}{m_i}\bigg)^{1-N_{if}/N_i}\,,
\label{Mivev}
\ee
with $m_i = c_i \phi^{q_i}$, being both suppressed by the dynamically generated scales $\Lambda_i\ll 1$ and the small ratio $\Lambda_i/m_i$.\footnote{Of course, we are assuming at this stage the existence of a non-runaway minimum for the moduli $U,Z$.} Correspondingly, we can completely neglect the mesons in $K_h$, so that the resulting effective K\"ahler potential is simply
\be
K_{eff}=K_M+\alpha_\phi \phi^\dagger e^{- 2V_X} \phi+ \alpha_\chi \chi^\dagger e^{ 2 q_\chi V_X} \chi \,.
\label{Keff}
\ee
As next step, we look for vacua with $\chi\ll 1$. We expand the scalar potential $V_{eff}$  arising from  (\ref{Weff}), (\ref{Keff}) and the $D_X$ term in powers of $\chi$, $V_{eff}=\sum_{n,m=0}^\infty V_{n,m} \chi^n\chi^{\dagger m}$, and keep only the leading term $V_0\equiv V_{0,0}$. It reads
\be
V_0=\frac{1}{2{\rm Re}\, f_X} (D_X^{(0)})^2+e^{K_{eff}^{(0)}} \bigg[ \sum_{i,j=U,Z,\phi} K_{eff}^{(0)i\bar j} F_i ^{(0)}F_{\bar j}^{(0)}
- 3 |W_{eff}^{(0)}|^2 + V_{up-lift}\bigg]
\,, \label{V0}
\ee
where the $F$-terms are computed using $K_{eff}$ and $W_{eff}$ and the superscript $(0)$ means that all expressions are evaluated for $\chi = 0$. The first three terms in $V_0$ correspond to the SUGRA scalar potential that would result from $K_{eff}^{(0)}$, $W_{eff}^{(0)}$ and $f_X$. The last term
\be
V_{up-lift}= \frac{| F_\chi^{(0)}|^2}{\alpha_\chi}\,,
\label{Vuplift}
\ee
where $F_\chi^{(0)} = f(\phi)\exp(- \gamma_Z Z -\gamma_U U )$, is effectively a  moduli-dependent  ``up-lifting'' term. 

Let us look for approximate SUSY vacua for $U$ and $Z$, neglecting for the moment the up-lifting term $V_{up-lift}$, and assuming that at the extremum $w_0$ is larger than the dynamically generated terms in (\ref{Weff}). It is easy to solve the system $D_X^{(0)}=F_U^{(0)} = F_Z^{(0)} = F_\phi^{(0)}=0$. The $D_X$ term trivially vanishes when eq.(\ref{phi0}) is satisfied, so that $\phi$ is determined.  At a SUSY extremum, gauge invariance implies $F_\phi^{(0)}=-\delta/(2\phi) F_U^{(0)}$, so that we are left to solve the system $F_U^{(0)} = F_Z^{(0)} = 0$.  We get
\be
U_{SUSY}  \simeq \frac{b_2 x_1 - b_1 x_2}{ a_1 b_2 - a_2 b_1}\,, \ \ \ \ \
Z_{SUSY} \simeq  \frac{a_1x_2 -a_2 x_1}{ a_1 b_2 - a_2 b_1}\,,
\label{UZsusy}
\ee
where 
\be
x_{1,2}   =  -\log\Bigg[\pm \frac{w_0}{A_{1,2}(\phi_{SUSY})}\frac{b_{2,1} K_{eff}^{(0)\prime}-a_{2,1} \dot K_{eff}^{(0)}}{a_1 b_2 - a_2 b_1}
\Bigg] \,,
\label{xycoeff} \ee and a dot stands for a derivative with respect to $Z$. By appropriately choosing the signs of $\eta_i$
appearing in $A_i(\phi)$, see eq.(\ref{abcoeff}), 
we can always set $U_{SUSY}$ and $Z_{SUSY}$ to be real, so that for simplicity of notation in the
following we will always assume real fields and real parameters. Since $U$ and $Z$ enter explicitly in the
coefficients $x_{1,2}$ above, eqs.(\ref{UZsusy}) do not admit explicit analytic solutions. However, the
logarithmic dependence on $U$ and $Z$ of $x_{1,2}$  is often mild enough that a good approximate expression for 
$U_{SUSY}$ and $Z_{SUSY}$ is obtained by taking some educated guess for the moduli in eq.(\ref{xycoeff}),
compute (\ref{UZsusy}), insert the result in (\ref{xycoeff}) and compute once again (\ref{UZsusy}). The shifts in the fields due to the up-lifting term $V_{up-lift}$ can be found by expanding
the extrema conditions $\partial_U V_0=\partial_Z V_0 = \partial_\phi V_0= 0$ around the SUSY values
(\ref{UZsusy}): $U=U_{SUSY}+\Delta U$, $Z=Z_{SUSY}+\Delta Z$, $\phi=\phi_{SUSY}+\Delta \phi$ and keeping the leading term in $V_{up-lift}$ and terms up to linear order in $\Delta U$, $\Delta Z$ or $\Delta \phi$ in the remaining terms of the scalar potential. The resulting expressions one gets for the shifts  are
actually very involved and can be handled only numerically. Some simple approximate formulae can however be
derived by using simple scaling  arguments to estimate the typical size of the terms entering in the K\"ahler and
superpotential terms, eqs.(\ref{Weff}) and (\ref{Keff}). We first notice that eq.(\ref{UZsusy}) fixes the sizes of
the moduli $U$ and $Z$ at the SUSY point to be inversely proportional to the effective parameters $a_{1,2}$ and
$b_{1,2}$. For simplicity, we can take all the $a_{1,2}$ and $b_{1,2}$ parameters that do not vanish to be of the
same order of magnitude and denote their common value by  $a$. It then follows that $U\sim Z$ and we can generally
denote by $X$ the common modulus VEV. We will use such simplified notation anytime we want to estimate a quantity
without giving its explicit expression. Coming back to eq.(\ref{UZsusy}), it is clear that $a X$ is approximately
a constant, proportional to the $x_{1,2}$ coefficients defined in eq.(\ref{xycoeff}). Since $w_0\ll 1$ is required
to have a sufficiently light gravitino mass, this constant is much larger than 1. 
As a matter of fact, for a wide class of models $aX$ is always in the narrow range $20 \lesssim a X \lesssim 40$, which 
is essentially the range dictated by the Planck/electroweak scale hierarchy.  
The parameter
$\epsilon\equiv 1/(a X)$ is then small and an expansion in $\epsilon$ is possible. 
The following scaling
behaviours are taken at the SUSY extremum:
\bea
\!\!\!\!&&   \partial_X^n K_{eff}^{(0)}\sim  \partial_X^n K_M\sim \frac{1}{X^n} \,, \ \hspace{2.9cm}
 \partial_X^n \alpha_{\phi,\chi} \sim\frac{\alpha_{\phi,\chi}}{X^n}\,, \ \ \ n>0\,.  \label{scalingX} \\
\!\!\!\! && \partial_X W_{eff}^{(0)} \sim (\partial_X K_{eff}^{(0)}) W_{eff}^{(0)}\sim  \frac{1}{X} W_{eff}^{(0)} \,,  \ \ \ \ \ \
 \partial_X^n W_{eff}^{(0)} \sim a^{n-1} \partial_X W_{eff}^{(0)}\,, \ \ \ n>1\,.  \nn
\eea 
The term proportional to $|\phi|^2$ in $K_{eff}^{(0)}$ is sub-leading in $\epsilon$ with respect to the purely
moduli dependent term $K_M$. Indeed,  $\alpha_\phi |\phi|^2 \lesssim \delta/X$ and, neglecting possible corrections due to $\eta_{iZ/U}$,  one has from eqs.(\ref{GSexplicit}) and (\ref{abcoeff}) that $\delta = 2q_i N_{if}/(N_i a_i)$, which typically
is less  or equal to $1/a$.
This explains the first relation in eq.(\ref{scalingX}).
We are now ready to estimate the shift of the fields $\Delta U$, $\Delta Z$ and $\Delta \phi$ due to the
up-lifting term $V_{up-lift}$. In the heavy $U(1)_X$ gauge field approximation $M_X\gg m_{3/2}$,  which will
always be the case of interest for us, one has (see e.g.\cite{GomezReino:2007qi}) 
\be
\langle D_X \rangle \simeq \frac{2}{M_X^2} e^{K_{eff}^{(0)}} \frac{q_\chi (F_\chi^{(0)})^2}{\alpha_\chi} = \frac{2}{M_X^2}q_\chi e^{K_{eff}^{(0)}} V_{up-lift} \,,
\label{Datvacuum} 
\ee 
where $M_X^2=  2 g_X^2 \alpha_\phi \phi^2_{SUSY} $,  so that the $D_X$ term at the minimum is negligibly small. We can use the condition
$D_X\simeq 0$ to express $\Delta \phi$ as a function of $\Delta Z$ and $\Delta U$.
Using eq.(\ref{phi0}), it is straightforward to see that $\Delta \phi$ scales as 
\be 
\Delta \phi \sim \phi \frac{\Delta X}{X}\,. 
\label{deltaphi}
\ee 
Next step is to estimate $\Delta X$. Using eqs.(\ref{phi0}), (\ref{scalingX}), (\ref{Datvacuum})  and (\ref{deltaphi}), it is a simple exercise to see that at leading order in $\epsilon$
and up to linear order in $\Delta X$, one has
\be
\partial_X V_0 \simeq e^{K_M^{(0)} }K_M^{X\bar X}\partial_{\bar X} F_{\bar X} \partial_X F_X \Delta X+\partial_X ( e^{K_M^{(0)} }V_{up-lift})+
e^{K_M^{(0)} }q_\chi V_{up-lift} \frac{\partial_X^2 K_M}{\partial_{X}K_M}
 \simeq 0\,,
\label{deltaXeq}
\ee
giving 
\be 
\Delta X \sim V_{up-lift}\frac{\partial_X^2 K_M \Big[q_\chi\partial_X^2 K_M+(\partial_X
K_M)^2 +\gamma \partial_X K_M\Big]}{{\partial_X K_M\Big(\partial_X^2 W_{eff}^{(0)}\Big)^2}} \sim \epsilon^2 \frac{V_{up-lift}}{(W_{eff}^{(0)})^2}
X (q_\chi+\gamma X)\,, \label{deltaX} 
\ee 
where $\gamma$ generally denotes $\gamma_U$ or $\gamma_Z$ and we have tacitly assumed $q_\chi >1$ in writing the last relation in eq.(\ref{deltaX}).
At a Minkowski minimum, $|V|_{up-lift}\sim |W_{eff}^{(0)}|^2$, so that the fraction in eq.(\ref{deltaX}) is
$O(1)$. If $\gamma_U=\gamma_Z=0$ and $q_\chi \sim O(1)$, the relative shifts of the moduli are small: $\Delta X/X \sim \epsilon^2\sim
10^{-3}$ and certainly the up-lifting term $V_{up-lift}$ does not de-stabilize the system and can be treated as a
perturbation, as we did. When $\gamma_U$ and/or $\gamma_Z$ are non-vanishing, the up-lifting term $V_{up-lift}$
becomes exponentially sensitive to the values of $U,Z$.  It is then not enough to have  $\Delta X/X \ll 1$, but
the stronger constraint  $\gamma \Delta X\ll 1$ is required, in order to avoid large displacements of
$V_{up-lift}$ which can result on the impossibility of finding  a Minkowski solution. This results on a
bound on the size of $\gamma$:
\be 
\gamma^2 \ll a^2\,. \label{gammabound} 
\ee
The same constraint  $\gamma \Delta X\ll 1$ gives also an upper bound on $q_\chi$, $\epsilon\, q_\chi \gamma/a\ll 1$, which is however 
quite mild, in light also of eq.(\ref{gammabound}).
We do not report the detailed expressions for $\Delta U$ and $\Delta Z$ in the general case, which are very involved and not illuminating even when expanded in powers of $\epsilon$. Just for concreteness, we report their form for the particular
case when $\alpha_\chi = \alpha_\phi = 1$, factorizable $K_M, W_{eff}$ (i.e. $\dot K_M^\prime = \dot W_{eff}^{\prime}=0$)
and perturbatively generated $Y$ Yukawa coupling: $\gamma_U = \gamma_Z = 0$. In these approximations, we have
\bea
\Delta U & \simeq &  - V_{up-lift} \frac{K_M^{\prime\prime} [(K_M^\prime)^2+q_\chi K_M^{\prime\prime}]}{K_M^\prime (W_{eff}^{(0)\prime\prime})^2} \,, \nn \\
\Delta Z & \simeq &  - V_{up-lift} \frac{\dot K_M \ddot K_M}{(\ddot W_{eff}^{(0)})^2} \,, \label{DeltaUZ} \eea
whose scalings are in agreement with the general estimate (\ref{deltaX}). Notice that both $\Delta Z$ and $\Delta U$
are positive, since $\dot K_M$ and $K_M^\prime$ are negative, tending to decrease the up-lifting term.
The scaling behaviours of the $F$ terms at the non-SUSY vacuum are easily found:
\bea
F_\chi^{(0)} & \sim & \alpha_\chi^{1/2} W_{eff}^{(0)}\,, \nn \\
F_X^{(0)} & \sim & \epsilon \Big(\gamma+\frac{q_\chi}{X}\Big)
\alpha_\chi^{-1/2} F_\chi^{(0)}\,,  \nn \\
 F_\phi^{(0)} & \sim &\frac{\delta}{\phi} \epsilon
  \Big(\gamma+\frac{q_\chi}{X}\Big)
 \alpha_\chi^{-1/2} F_\chi^{(0)}\,.
 \label{FtermsSca}
\eea
Using eq.(\ref{FtermsSca}), one can easily estimate $K_{eff}^{(0)i\bar j} F_i ^{(0)}F_{\bar j}^{(0)}$ and $V_{up-lift}$.  In agreement with our expectation,  $V_{up-lift}$ is the leading term contributing positively to the vacuum energy, justifying its name of up-lifting term.
It is also possible to verify the validity of our approximation of having integrated out the meson fields $M_i$.
Having integrated them out in the flat-space limit, the $F_{M_i}$ at the non-SUSY vacuum scale as in \cite{FIterm}
and are of order $F_{M_i} \sim W_{eff}^{(0)}$. However, they are totally negligible due to powers of $M_i$ which
appear in the K\"ahler metric, i.e. $F^{M_i} \sim M_i F_{M_i}\ll F_{M_i}$.\footnote{The meson masses induced by the VEV's
of $\phi$ and the moduli are typically much higher than the dynamically generated scales, so that one can safely
take the classical K\"ahler potential for the mesons: $K(M_i) = 2{\rm Tr}(M_i^\dagger M_i)^{1/2}$. Equivalently, the description
in terms of quarks is valid at these energy scales.}

Once the approximate vacuum of the leading potential $V_0$ has been found,  given by
$U_0=U_{SUSY}+\Delta U$, $Z_0=Z_{SUSY}+\Delta Z$,  $\phi_0=\phi_{SUSY}+\Delta \phi$, 
we turn on $\chi$ and verify the validity of our initial assumption $\chi\ll 1$.
The easiest way to estimate $\chi$ is to use the second relation in eq.(\ref{FDrel}) that relates the $D_X$ term to the $F$-terms.
At the minimum $D_X\simeq 0$, but the $F$-terms for $\chi$ and $\phi$ contributing to $D_X$ are typically of the same order of magnitude: $\phi F_\phi \sim q_\chi \chi F_\chi $. Using this relation and the scalings (\ref{FtermsSca}), we get
\be
\chi_0 \sim \delta \epsilon \Big( \gamma+ \frac{q_\chi}{X}\Big)\frac{1}{q_\chi} \alpha_\chi^{-1/2}\,,
\label{chiscal}
\ee
which proves our initial assumption $\chi\ll 1$.
A more accurate estimate of $\chi_0$ might be obtained by considering the next sub-leading potential terms $V_{1,0}=V_{0,1}$, $V_{2,0}=V_{0,2}$ and $V_{1,1}$ obtained by expanding $V_{eff}$ in powers of $\chi$ and $\chi^\dagger$. In first approximation, one can freeze 
$U$, $Z$ and $\phi$ at the values $U_0$, $Z_0$ and $\phi_0$, which extremize $V_0$, so that  $\chi$ is determined by a linear equation.
The general explicit expression for $\chi$ is however very involved and not very interesting, so we will not report it here.
For the same reason, we do not report the expressions of the further shifts of $U$, $Z$ and $\phi$ induced by the backreaction of $\chi$.
They turn out to scale as eqs. (\ref{deltaphi}) and
(\ref{deltaX}), but are typically smaller in the parameter region we will consider in the following. The location of the vacuum is then slightly shifted but it is not destabilized by the field $\chi$. We can also check how $\chi$ changes the values of the $F$-terms 
(\ref{FtermsSca}). One has 
\bea
F_\chi & =  &  F_\chi^{(0)}+\alpha_\chi \chi (W_{eff}^{(0)}+\chi F_\chi^{(0)}),\nn \\
F_X & = & F_X^{(0)}+\chi (\partial_X F_\chi^{(0)}+\partial_X K_{eff} F_\chi^{(0)}),\nn \\  
F_\phi&  = &  F_\phi^{(0)}+\chi \Big( \frac{\hat q_\chi}{\phi} +\phi \,\alpha_\phi \Big)  F_\chi^{(0)}. 
\label{fullFterms}
\eea
Using eqs.(\ref{FtermsSca}) and (\ref{chiscal}), it is straightforward to verify that the effect of $\chi$ on $F_\chi$ and $F_X$ is negligible, while the second term in $F_\phi$  is of the same order as $F_\phi^{(0)}$. Hence the scalings (\ref{FtermsSca}) hold also for the
full $F$-terms $F_\chi$, $F_X$ and $F_\phi$, providing a final consistency check of eq.(\ref{chiscal}), which has been derived
under this assumption. 

Let us now discuss under which conditions the above SUSY breaking mechanism is stable under small deformations.
The choice of the superpotential (\ref{fullWuv}) was rather ad hoc, since we have considered only linear terms in $\chi$ and tacitly assumed that possible higher order terms of the form $Y_n(U,Z,\phi) (\chi \phi^{q_\chi})^n$, with $n>1$, can be neglected.
This assumption is actually very strong, since the requirement $|F_\chi^{(0)}|\simeq \alpha_\chi^{1/2} |W_{eff}^{(0)}|$ 
puts severe constraints on the size of the constant term $Y$ appearing in (\ref{yukawauv}). This is particularly clear
if one notices that generally $\alpha_\chi \lesssim 1$ and $W_{eff}^{(0)}\simeq w_0\lesssim O(10^{-13})$.
The more obvious options of assuming a perturbative ($\gamma_{U,Z}=0$) mass term  ($q_\chi=1$) or trilinear coupling  ($q_\chi=2$) 
leads to an unnaturally small coupling $Y$. In such a situation, the terms of the form $Y_n(U,Z,\phi) (\chi \phi^{q_\chi})^n$  will lead to a restoration of SUSY and to the destabilization of the non-SUSY vacuum. This is best seen by considering the flat space model with stabilized moduli.
In this case, the relevant superpotential term in eq.(\ref{Weff}) is just the term linear in $\chi$, which is
invariant under a $U(1)_R$ symmetry with $R(\chi)=2$, $R(\phi)=0$. An exact $R$-symmetry is generally necessary to get a SUSY-breaking vacuum \cite{Nelson:1993nf} and, indeed, in absence of moduli and gravitational dynamics, $\chi$ is stabilized at the origin where $U(1)_R$ is unbroken. Any term of higher order in $\chi$ will necessarily break $U(1)_R$, leading to the appearance of
SUSY vacua. Gravity and moduli explicitly break $U(1)_R$, but if the breaking is small enough
their only effect would be to displace a bit $\chi$ from the origin, so that $\chi\ll 1$, as predicted by eq.(\ref{chiscal}). If the terms  $Y_n(U,Z,\phi) (\chi \phi^{q_\chi})^n$ are all negligibly small, much smaller than $\chi w_0$, the SUSY preserving vacua will appear for large values of $\chi$ and will not perturb much the
(meta)stable non-SUSY vacuum close to the origin. However, when the terms $Y_n(U,Z,\phi) (\chi \phi^{q_\chi})^n$ become roughly of the same order as $\chi w_0$, the SUSY vacua approach the origin and the non-SUSY
vacua are destabilized and disappear.

Invoking non-perturbatively generated couplings ($\gamma\neq0)$ alleviate the problem, but it does not solve it, 
because a natural $Y$ would require $\gamma\sim a$,  so that
\be e^{-\gamma \,U}\sim e^{-a \, U}\sim W_{eff}^{(0)}\,,
\ee
but the constraint (\ref{gammabound}) does not allow such values of $\gamma$ . 
A possible way out is to consider higher-order couplings by taking $q_\chi >2$, so that
the effective term $\phi_0^{q_\chi}$ becomes small enough to get not so small values of $Y$. In this way, we also more effectively
suppress the dangerous terms  $(\chi \phi^{q_\chi})^n$. Summarizing, the requirement of naturalness and more importantly
stability under superpotential deformations with higher powers of $\chi$ necessarily require to consider non-renormalizable
interactions with $q_\chi >2$, the precise bound on $q_\chi$ depending on $\gamma_{U,Z}$ being zero or not.\footnote{A simple way to overcome this problem is to assume that $Y$ is an effective non-perturbatively generated coupling of some other modulus which has been already stabilized. 
This explanation, however, requires an effective decoupling between the stabilized modulus 
and the sector of the theory responsible for SUSY breaking.}

So far, we have been able to find approximate expressions for the extrema of the scalar potential $V_{eff}$, but
we have still to check whether these vacua are minima or not. 
At leading order in $\epsilon$ and for $\phi_{SUSY}/X\ll 1$, the kinetic mixing of $\phi$ with the moduli can be neglected
and the mass of $\phi$ is determined by the D-term potential. Its physical mass is 
\be 
m^2_\phi \simeq 2 g_X^2 \alpha_\phi \phi^2_{SUSY} \,, \label{MPHI}
\ee 
which is also the mass of the gauge vector boson $A_X$, as we have seen. Indeed, ${\rm Im}\,\phi$ is  approximately
the would-be Goldstone boson eaten up by $A_X$ after the $U(1)_X$ gauge symmetry breaking.
The leading contribution to the $\chi$ mass is also easily derived by looking at the terms quadratic in $\chi,\chi^\dagger$.
Its physical mass equals
\be m_\chi^2
\simeq \frac{2q_\chi^2}{\alpha_\phi \phi_{SUSY}^2}V_{up-lift}\,.\label{MCHI} 
\ee
From eq.(\ref{MCHI}) we have $m_\chi\gtrsim (q_\chi/\sqrt{\epsilon})\, m_{3/2}\gg m_{3/2}$ and hence the effects of SUSY breaking
on $m_\chi$ are small, so that $m_\chi$  is approximately the mass of both components of the complex field $\chi$.
The mass scale of $m_\phi$ is of order $\sqrt{\epsilon /X}$ and, unless the moduli are very large, it is 
just one or two orders of magnitude below the Planck scale. The moduli masses are more involved and are best described by an effective approach where the massive fields $\phi$ and $A_X$ are integrated out, approach  we will now consider.

\subsection{Effective Description}

As we have seen, the cancellation of the $U(1)_X$ D-term given by a non-trivial VEV for $\phi$ induces large meson masses,
as well as large masses for $A_X$ and $\phi$ itself. One might then not only integrate out the meson fields, as we did,
but also $A_X$ and $\phi$. In first approximation,  we can integrate out $\phi$ and $A_X$ at the SUSY level.
A useful way to do that is to go to the super-unitary gauge, which is the super-field version of the standard unitary gauge. By keeping the first terms in an expansion in $\epsilon$, the super-unitary gauge reads 
\be
\phi = \phi_{0} - \frac{1}{2\alpha_\phi \phi_0} \bigg[(\delta K_M^{\prime\prime} + 2\phi_0^2 \alpha_\phi^\prime) (U-U_0)+ (\delta \dot K_M^{\prime}+ 2\phi_0^2 \dot\alpha_\phi) (Z-Z_0) \bigg]
\,, \label{superunitary} 
\ee
where $\phi_0$, $U_0$ and $Z_0$ are the approximate VEV's we have previously found.
The would-be Goldstone boson is essentially given by ${\rm Im}\,\phi$, being the last two terms in eq.(\ref{superunitary}) suppressed at least by a factor $\sqrt{\epsilon}$. 
In this way, we can get rid of the $\phi$ chiral field, substituting eq.(\ref{superunitary}) (and its anti-chiral
version) in both the K\"ahler potential (\ref{Keff}) and superpotential (\ref{Weff}). At leading order, we can neglect the
last two terms in eq.(\ref{superunitary}) and just take $\phi= \phi_0$. 
We can then expand the resulting effective K\"ahler potential $K_{eff}(\phi_0)$ for small $V_X$ and keep up to quadratic terms in $V_X$. 
Then the equation of motion for $V_X$, $\partial K_{eff}(\phi_0)/\partial V_X=0$ is easily solved and one finds
\be
V_X  \simeq -\frac{D_X}{2\alpha_\phi \phi_0^2} \simeq -\frac{ q_\chi \alpha_\chi }{2\alpha_\phi \phi_0^2}|\chi|^2\,,
\label{VxVEV}
 \ee
where in the last relation we have completely neglected the $U$ and $Z$ dynamics by
taking $D_X\simeq q_\chi \alpha_\chi |\chi|^2$. 
By plugging back eq.(\ref{VxVEV}) in $K_{eff}(\phi_0)$, we get the following effective K\"ahler potential at leading 
order \cite{NAHetal}:
\be \hat K_{eff}\simeq  \alpha_\chi |\chi|^2  + K_M -\frac{D_X^2}{2\alpha_\phi \phi_0^2}
\simeq \alpha_\chi  |\chi|^2 +K_M -\frac{q_\chi^2 \alpha_\chi^2}{2\alpha_\phi \phi_0^2}  |\chi|^4 \,.
\label{Keff2}
\ee
The superpotential $\hat W_{eff}$ trivially follows from (\ref{Weff}) with $\phi=\phi_0$, so that the field dependent terms $f(\phi)$ and $A(\phi)$ defined in eq.(\ref{abcoeff}) become now effective constants $f(\phi_{0})$ and $A(\phi_0)$.
Notice that $\hat K_{eff}$ and $\hat W_{eff}$ sensitively depend on $\phi_0$, whose precise value cannot be correctly determined without actually using the full underlying model. However, the shifts on $\phi$, as computed in the previous subsection, are small enough that
at leading order one might safely replace $\phi_0$ in all the above formulae (and the one that will follow) by the SUSY value (\ref{phi0}).

The effective model described by  $\hat K_{eff}$ and $\hat W_{eff}$ is considerably more tractable than the underlying UV model
we considered before. In particular, some physical features are more transparent and, in addition, such effective description provides us with an approximate formula for the moduli masses. For instance,  it is immediately clear  that a vacuum with
$\chi_0 \ll 1$ and non-runaway moduli will necessarily break SUSY, since $F_\chi \simeq f \exp(-\gamma_Z Z_0-\gamma_U U_0)\neq 0$. In fact, as far as $\chi$ is concerned, the model is nothing else than a Polonyi model with a deformed K\"ahler potential. As we already mentioned,
in the flat-space limit with decoupled moduli, $\chi$ will be stabilized at the origin due to the $|\chi|^4$ term in eq.(\ref{Keff2}).
The extrema conditions (\ref{UZsusy}) for $U$ and $Z$ are rederived by requiring $\hat F_U^{(0)}=\hat F_Z^{(0)}=0$, using the same notation as before. An approximate analytical formula for the moduli mass terms can be derived, once again
by keeping the leading term in an expansion in $\epsilon$. One gets
\be
m^2_{i\bar j} \simeq e^{K_M} K_M^{\bar ml} \partial_i \partial_l \hat W_{eff}^{(0)} \partial_{\bar j} \partial_{\bar m} \bar{\hat W}_{eff}^{(0)} \,,
\ee
with the indices running over $U$ and $Z$. Since $K_M$ is a function of the real part of the moduli only, we can drop any distinction between holomorphic and anti-holomorphic indices in taking derivatives of the K\"ahler potential $K_M$. The K\"ahler metric $g$ for the moduli is then a real symmetric matrix which is diagonalized by an orthogonal $SO(2)$ matrix $C$: $C^t g C = d$, with $d$ a diagonal matrix. It follows that the physical moduli masses are given by the eigenvalues of the following mass matrix:
\be
d^{-1/2} C m^2 C^t d^{-1/2}\,.
\label{massmatrix}
\ee
In order to have a minimum, we have to require that the mass matrix (\ref{massmatrix})
is positive definite. This will in general give a non-trivial constraint on the possible form of $K_M$ and
the parameter space of the model. However, such a constraint is not very restrictive, as can be seen by considering, 
for instance, the particular case of decoupled K\"ahler and superpotential terms, namely $\dot {\hat K}_M^\prime = \dot{\hat  W}_{eff }^\prime=0$. In this simple case, both the moduli 
masses and kinetic terms are already in a diagonal form, and we get 
\be
m_{U,Z}^2 \simeq  e^{K_M} \left| \frac{\partial_{U,Z}^2 \hat W_{eff}^{(0)}}{\partial_{U,Z}^2 K_M}\right|^2 \sim \epsilon^{-2} m_{3/2}^2\,,
\label{MassModuli}
\ee
which is manifestly positive definite.
The last equality of eq.(\ref{MassModuli}) shows the scaling of the moduli masses with respect to the gravitino mass $m_{3/2}$. As can be seen, the moduli are parametrically heavier than the gravitino, which is a cosmologically welcome feature.

An important comment is now in order.
One might wonder why we have decided to adopt from the beginning an effective field theory approach  for the hidden mesons (and tacitly for all the other moduli possibly responsible for the constant term in the superpotential), integrating
them out from the very beginning, and not for $\phi$ and $A_X$ which are actually even heavier !
From a purely effective quantum field theory point of view, this is indeed not justified, the correct procedure being the integration
of all the states in the order specified by their mass scales and run the effective parameters down to lower energies.
At the classical level we are considering here and when focusing only on the properties of the vacua,
however, no real difference occurs and integrating out some state or not is only a matter of simplicity.
Contrary to the mesons, which can always be easily integrated out supersymmetrically to a very good approximation,  as we did,
$\phi$ and $A_X$ would require more work than what we have shown above to be properly integrated out, because they are more
sensible to SUSY breaking effects. If one wants to go beyond the leading terms in $\epsilon$ and study more quantitatively the model, in particular, we should keep the full super-unitary gauge (\ref{superunitary}) instead of taking $\phi = \phi_0$. More importantly, the approximation of completely neglecting the moduli dynamics in $D_X$ and substitute it with just $q_\chi \alpha_\chi |\chi|^2$, as we did in eqs.(\ref{VxVEV}) and (\ref{Keff2}), turns out to be in general a quite crude approximation.
Both these approximations can be relaxed and we have analytically and numerically checked that 
the resulting ``improved" effective model reproduce pretty well, at a more quantitative level, the main properties of the full theory. 
Contrary to the naive effective theory we have shown above, however, the improved theory is no less complicated than the
full one, so that no real simplification occurs in considering it. On the contrary, the naive model captures
all the qualitative features of the full model and, as a matter of fact, it has been crucial to guide us in the  
analysis of the previous subsection.

The analysis of the general two-moduli model performed so far can trivially be reduced to the single modulus ($U$) case
with essentially no effort when $w_0\neq 0$. We will then not repeat the analysis here, but just point out that all the considerations
we made for the two-moduli case apply.
The only qualitative difference is that with a single modulus one condensing gauge group is enough to get (meta)stable vacua. In particular, we have reproduced within our approach the results of \cite{Dudas:2007nz} for a KKLT--like model with a single modulus.

\section{Models with $w_0=0$}

From the previous discussion it is clear that in general one has to face a possible fine-tuning problem to get $Y\ll 1$
in the Yukawa sector for $\chi$, but one might argue that this is, by means of the flatness condition,  just a reflection of the 
other fine-tuning problem required to have a tiny constant superpotential term, $w_0\ll 1$. In principle, then, we have to face up to three
fine-tuning problems, the third being the unavoidable tuning of the cosmological constant.
Moreover, up to possible suppressions coming from the $\exp(K)$ term in the scalar potential, $w_0$ essentially fixes the 
supersymmetry breaking scale. It would be more desirable, instead, to dynamically generate it. 
This motivates us to analyze also the case in which $w_0$ vanishes, assuming that some stringy symmetry
forbids its appearance. Let us start by considering a theory with a single modulus ($U$).
Most of the considerations we made for $w_0\neq 0$ continue to apply for $w_0=0$,
the main difference being the moduli stabilization mechanism, which now  boils down to a racetrack model \cite{Krasnikov:1987jj}, 
where the scale of supersymmetry breaking is dynamically generated.  
The effective K\"ahler potential is still given by eq.(\ref{Keff}), with the obvious understanding that  $K_M$, $\alpha_\phi$ and
$\alpha_\chi$ do not depend on $Z$. Similarly, the effective superpotential is as in eq.(\ref{Weff}), with $\gamma_Z = b_1= b_2=w_0=0$.
The condition of vanishing $D_X$ term still fixes $\phi$ to the value (\ref{phi0}). The equation $F_U^{(0)}=0$, at leading order in $\epsilon$  has as solution
\be
U_{SUSY}=\frac{1}{a_1-a_2} \ln\left(-\frac{a_1 A_1(\phi_{SUSY})}{a_2 A_2(\phi_{SUSY})}\right)\,.
\label{Ususyw0}
\ee
The axionic component of $U$ is always extremized such that the two condensing sectors get opposite signs, therefore for simplicity we take $\eta_1 = -\eta_2 = 1$ and set it to zero.
The scaling relations reported in eq.\eqref{scalingX} are still valid but the very last relation among the derivatives of the superpotential should be reviewed. Indeed, the racetrack models work using the competing effects of the different condensing sectors, as clearly illustrated by eq.(\ref{Ususyw0}). As a result, there
are some cancellations among the condensing scales that hold at the
$F$-term level, but are destroyed once derivatives of $F$-terms are taken. The result of this is that the scaling behavior of
the first derivative of the superpotential still satisfies the relation given by 
\eqref{scalingX}, but the higher ones are changed to 
\be
\partial_U^n W_{eff}^{(0)} \sim  a^{n}  W_{eff}^{(0)}\,, \ \ \ n>1\,.
\label{newscalingW}
\ee
Eq.(\ref{deltaXeq}) and the first relation in eq.(\ref{deltaX})  still hold with obvious notation changes, 
so using eq.(\ref{newscalingW}),we now get for the modulus shift:
\be 
\Delta U \sim  \epsilon^4 \frac{V_{up-lift}}{(W_{eff}^{(0)})^2}\,
U(q_\chi+\gamma\, U)\,, \label{deltaXRT} 
\ee
where we omit the unnecessary subscript $U$ in $\gamma$.
Comparing eq.(\ref{deltaXRT}) with eq.(\ref{deltaX}), we notice that the shift of $U$ induced by the up-lifting term
is now two or three orders of magnitude smaller than the shift in the models with $w_0\neq 0$,  being $O(\epsilon^4)$ instead of 
$O(\epsilon^2)$. The $F$-terms are also parametrically smaller than before:
\be F_U^{(0)} \sim \epsilon^2\, \Big(\gamma+\frac{q_\chi}{U}\Big)
\alpha_\chi^{-1/2} F_\chi^{(0)}\,,  \ \ \ \ \ 
F_\phi^{(0)}  \sim  \frac{\delta}{\phi} \epsilon^{2}
  \Big(\gamma+\frac{q_\chi}{U}\Big)
  \alpha_\chi^{-1/2} F_\chi^{(0)}\,,
\label{RTFterm}
\ee 
where $F_\chi^{(0)}  \sim  \alpha_\chi^{1/2} W_{eff}^{(0)}$.
The constraint $\gamma \Delta U \ll 1$ gives now
\be
\gamma^2\ll  \frac{a^2}{\epsilon^2}\,.
\label{gammaboundRT}
\ee
We see from eq.(\ref{gammaboundRT}) that  values of $\gamma \sim a $ are now allowed, 
solving the fine-tuning problem in the coupling $Y$. 
The shift on the vacuum induced by the backreaction of $\chi$ is now comparable or even slightly larger than eq.(\ref{deltaXRT}). However, it is typically small enough not to destabilize the vacuum. The scaling of $\chi$ can still be estimated by
the relation $\phi F_\phi \sim q_\chi \chi F_\chi$, giving
\be
\chi_0 \sim \delta \epsilon^2 \Big( \gamma+ \frac{q_\chi}{X}\Big)\frac{1}{q_\chi} \alpha_\chi^{-1/2}\,,
\label{chiw0}
\ee
which is small, as required. The scalings of the full $F$ terms (\ref{fullFterms}) can easily be worked out using eq.(\ref{chiw0}).
We find that $F_\chi\sim F_\chi^{(0)}$, $F_X\sim F_X^{(0)}$ and $F_\phi\sim F_\phi^{(0)}$, but the $\chi$-dependent terms
in $F_X$ and $F_\phi$ are non-negligible. The mass of $U$ can be estimated using an effective description, as explained in subsection 2.2. The first relation in eq.(\ref{MassModuli}) still holds, but the different scaling (\ref{newscalingW}) 
of the superpotential gives now
\be
m_U^2 \sim \epsilon^{-4} m_{3/2}^2\,.
\label{ModuliMassesw0}
\ee

The above analysis can be extended to the case of two moduli, in which case
one has to work with at least three condensing sectors to get viable SUSY solutions. 
Instead of considering the most general model with three condensing gauge groups, we will now
focus on an interesting class of models with decoupled non-perturbative superpotential terms. The effective superpotential $W_{RT3}$ reads now
\be
W_{RT3} =  f(\phi) e^{- \gamma_Z Z-\gamma_U U } \chi +A_1(\phi) e^{-a_1 U} - A_2(\phi) e^{-a_2 U} + A_3(\phi) e^{-b Z} \,.
\label{WRT3}
\ee
If the condensing scales associated to the gauge groups $G_1$ and $G_2$ are much larger
than that of $G_3$, $U$ is approximately stabilized by a racetrack mechanism given by $G_1$ and $G_2$ at the value (\ref{Ususyw0}). With $U$ so stabilized, $W_{RT3}^{(0)}$ (notation as before) 
boils down to a KKLT-like superpotential which gives the following SUSY extremum for $Z$ (see e.g. \cite{Serone:2007sv} for a similar analysis):
\be
Z_{SUSY}\simeq - \frac{1}{b} \log\bigg[\frac{\dot K_M \hat w_0}{b A_3}\bigg]\,,
\label{Zsusyw0}
\ee
where 
\be
\hat w_0\equiv A_1(\phi_{SUSY}) e^{-a_1 U_{SUSY}} - A_2(\phi_{SUSY}) e^{-a_2 U_{SUSY}}
\ee
is an effective constant superpotential term. 
The shifts in the fields induced by the up-lifting term $V_{up-lift}$ can be derived using the by now familiar expansion in $\epsilon \simeq 1/(b Z) \simeq 1/(a_1 U)\simeq 1/(a_2 U) \sim 1/(a U)$.
The form of the superpotential and the corresponding different stabilization mechanisms for $Z$ and $U$ do not allow now to consider $U$ and $Z$ together. Indeed, we have now
$\partial_U^n W_{RT3}^{(0)} \sim  a^{n}  W_{RT3}^{(0)}$, as in eq.(\ref{newscalingW}), and $\partial_Z^n W_{RT3}^{(0)} \sim  b^{n-1}  W_{RT3}^{(0)}/Z$, as in eq.(\ref{scalingX}). The leading terms in the shifts of the fields are as follows:
\be
\Delta Z \sim \epsilon^2 q_\chi X + \epsilon^2 X^2  \gamma_Z  + \epsilon^3 X^2  \gamma_U \,, \ \ \ \ \  
\Delta U\sim \epsilon^3 q_\chi X + \epsilon^3 X^2  \gamma_Z  + \epsilon^4 X^2  \gamma_U \,,
\ee
where $X$ denotes generically $U$ or $Z$, assumed to be of the same order of magnitude.
The usual bound $\gamma_Z \Delta Z \ll 1$ does not allow for natural values $\gamma_Z \sim a \sim b$, so that we are forced to consider $\gamma_Z=0$ and $\gamma_U\neq 0$, in which case 
$\gamma_U \sim a \sim b$ can be taken. The shift of $\phi$ is given by
$\Delta \phi \sim \phi \Delta  Z/Z$. The $F^{(0)}$-terms scale  as in eq.(\ref{RTFterm}), with 
$F_Z^{(0)}\sim F_U^{(0)}$. The considerations made for $\chi_0$ and the full $F$-terms
in the single modulus case apply also here.  The moduli masses depend on the form of the K\"ahler potential $K_M$.
If $K_M$ is factorizable, then $U$ and $Z$ will have masses roughly given by eqs.(\ref{ModuliMassesw0}) and (\ref{MassModuli}), respectively.
If $K_M$ is not factorizable, then generally the mass of $U$ will be essentially as given by eq.(\ref{ModuliMassesw0}), whereas $Z$ will be heavier than what predicted by eq.(\ref{MassModuli}), depending on the mixing between the two moduli in $K_M$.

\section{Soft Masses}

The natural framework of SUSY breaking mediation in any model with hidden and not sequestered
sector, is gravity mediation, with $m_{3/2}\sim O({\rm TeV})$.
A sufficiently heavy gravitino is desirable for cosmological reasons, being the moduli masses proportional to $m_{3/2}$, according to eqs.(\ref{MassModuli}) and (\ref{ModuliMassesw0}).\footnote{One might further push $m_{3/2}$ to O(10 TeV) or more, assuming a sequestering of the hidden sector from the visible sector, so that the gravity mediation can be suppressed and anomaly mediation takes over \cite{Randall:1998uk}. We will not consider this possibility, which is non-generic.}
When the would-be anomalous $U(1)_X$ gauge field is very massive, like the scenario advocated in our paper, one
can effectively integrate $A_X$ out and get the effective K\"ahler potential (\ref{Keff2}), taking care of including
possible visible sector contributions to the $D_X$ term, which we will shortly discuss.
In this way, all soft parameters can be derived using the standard results of \cite{Kaplunovsky:1993rd} with $F$-term breaking only. 
We will not explicitly compute all the resulting soft terms, but rather
we will just estimate the size of the gaugino and scalar masses. 

Let us start by considering the non-holomorphic soft scalar masses, in which case our considerations will apply to both the models
with $w_0\neq 0$ and $w_0=0$. For canonically normalized fields, they read \cite{Kaplunovsky:1993rd}:
\be
m_{i\bar i}^2 = m_{3/2}^2 - \frac{1}{\alpha_{iv}}F^I F^{\bar J} R_{i\bar i I\bar J}\,.
\label{generalsoftmass}
\ee
In eq.(\ref{generalsoftmass}), $F^I = \exp(\hat K/2)\hat K^{I\bar J}F_{\bar J}$, $I,J$ run over the hidden sector fields ($U$, $Z$ and $\chi$) and $\alpha_{iv}$ is the moduli-dependent function appearing in the K\"ahler potential of the visible sector, eq.(\ref{Kahlervisible}),
with $i$ running over the visible sector fields. Two possibilities arise, depending on whether the $U(1)_X$ charges $q_{iv}$ are vanishing or not.  When $q_{iv}\neq 0$,  $D_X\simeq \alpha_\chi q_\chi |\chi|^2 + \alpha_{iv} q_{iv} |Q^{(iv)}|^2$ and the leading canonically normalized soft mass terms  read
\be
m^2_{i\bar j} \simeq \delta_{i\bar j} m_{3/2}^2 \frac{3 q_\chi q_{iv}}{\alpha_\phi |\phi_0|^2 }\gtrsim  \delta_{i\bar j}  m_{3/2}^2 \frac{q_\chi q_{iv}}{\epsilon}\,,
\label{scalarsoftmass}
\ee
which arise from the term $F^\chi F^{\bar \chi} R_{\chi\bar \chi i \bar j}$ in eq.(\ref{generalsoftmass}). Using eq.(\ref{Datvacuum}), eq.(\ref{scalarsoftmass}) can be also rewritten in the more conventional form $m^2 = q_{iv} g_X^2 \langle D_X \rangle$. 

If $q_{iv}= 0$, $D_X\simeq  \alpha_\chi q_\chi |\chi|^2 -\delta/2 \alpha_{iv}^\prime q_{iv} |Q^{(iv)}|^2$. The leading term coming from $D_X$ is now of the same order as the universal $m_{3/2}^2$
term appearing in eq.(\ref{generalsoftmass}), so that 
\be
m^2_{i\bar j} \sim \delta_{i\bar j} m_{3/2}^2 \,.
\label{scalarsoftmassq0}
\ee
In eq.(\ref{scalarsoftmassq0}) we have not considered the contribution of possible quartic terms in the charged 
fields of the form $|Q_{iv}|^2 |\chi|^2$ which we have not specified in the K\"ahler potentials (\ref{Kahlervisible}) and (\ref{fullKuv}).
Their contribution can be relevant or even dominant, but it is model-dependent and can easily be derived from eq.(\ref{generalsoftmass})
once these terms are specified.
Given eqs.(\ref{scalarsoftmass}) and (\ref{scalarsoftmassq0}), the choice $q_{iv}=0$ is preferred, giving rise to not too heavy scalar masses.

Let us now consider the gauginos. Their canonically normalized masses are 
\be
m_g =\Big| F^I \frac{\partial_I f_{v}}{2{\rm Re}\, f_v}\Big|\,,
\label{generalgaugino}
\ee
where $f_v$ schematically denotes the holomorphic gauge kinetic functions of the visible gauge group.
Let us first discuss the models with $w_0\neq 0$. In this case, 
using eq.(\ref{FtermsSca}), we can easily estimate, for linearly moduli dependent $f_v$, 
\be
m_g \sim m_{3/2}\,  \epsilon\, X \Big(\gamma+\frac{q_\chi}{X}\Big)\,.
\label{gauginow0}
\ee 
We see that $m_g<m_{3/2}$, but on the other hand eq.(\ref{gauginow0}) predicts gaugino masses which are typically
larger than those found in the original KKLT scenario with $\bar D_3$ brane(s). This was already observed in \cite{Dudas:2007nz}
for a model with one modulus and perturbative up-lifting term ($\gamma=0$). We notice here that when $\gamma\neq 0$ (or $q_\chi>1$),
eq.(\ref{gauginow0}) predicts even larger gaugino masses.
In fact, considering that $\epsilon X \gamma \sim \gamma/a$ and the bound (\ref{gammabound}), the gauginos
can be made just a few times lighter than $m_{3/2}$, sufficiently heavy to neglect anomaly mediation contributions which 
become relevant if $m_g \lesssim m_{3/2}/(4\pi)$. We believe this is an important welcome feature of models with two moduli.
As already argued in \cite{Dudas:2007nz}, in presence of one modulus only, non-vanishing tree-level gaugino masses would require
$f_v$ to depend on $U$. Due to the non-linear transformation of $U$ under $U(1)_X$,  anomalous transformations of the action
are induced, which must be compensated by $U(1)_X$-$G_{vis}^2$ anomalies in the fermion spectrum, requiring $q_{iv}\neq 0$ or some
other modification, such as the introduction of  $U(1)_X$ charged fields, vector-like with respect to $G_{vis}$, which can also
be seen as messenger fields of a high scale gauge mediation. This possibility --- that would require
to study the backreaction of the messengers on our vacuum --- has been proposed in \cite{Dudas:2007nz} to alleviate the
hierarchy between the scalar and gaugino 
masses. We simply notice that in presence of two moduli, a more economical choice is to assume $f_v$ to depend on the neutral modulus $Z$ only, in which case one can safely take $q_{iv}=0$. We will see in a specific example in the next section that this choice, together with $\gamma_Z\neq 0$, gives rise to a fully satisfactory scenario for gaugino and scalar mass terms.

The gaugino masses in the models with $w_0=0$ sensitively depend on how we choose the exponential term $\gamma$
in the up-lifting term. As we have seen, a natural up-lifting term requires $\gamma_U \neq 0$ and hence $\gamma_Z=0$ if
we allow the non-perturbatively generated coupling to depend on one modulus only.
Furthermore, if we want to avoid introducing additional $U(1)_X$ charged fields, then $f_v$ should depend on $Z$ only.
In this case eq.(\ref{RTFterm}) gives, for linearly moduli dependent $f_v$:
\be
m_g \sim m_{3/2}\, \epsilon^2 \, X\Big(\gamma_U+\frac{q_\chi}{X}\Big)\sim m_{3/2}\,\epsilon \,,
\label{gauginonow0}
\ee
where in the last relation the scaling $\gamma_U\sim a$ has been used.
The gaugino masses are significantly lighter than the gravitino now, so that anomaly mediated
contributions cannot be neglected. Gaugino masses  can be increased by allowing $\gamma_Z$ to be non-vanishing, in which case they scale as in eq.(\ref{gauginow0}).  If one allows the non-perturbatively generated up-lifting term to depend on both moduli,
then  $\gamma_U$ and $\gamma_Z$ can respectively solve the naturalness problem of the up-lifting coupling
and alleviate the modest hierarchy between gaugino and scalar masses. 

\section{Explicit Models}

The exponential sensitivity of the superpotential  (\ref{Weff}) on the moduli, the not so small value  of the expansion parameter 
$\epsilon \sim 1/30$ and the several other approximations made before do not generally allow for a reliable, quantitative analytical study of the theory. Indeed, the main aim of sections 2 and 3 was to qualitatively characterize the models and to show the existence of metastable Minkowski vacua with low-energy SUSY breaking in a large area in parameter and moduli space, rather than quantitatively study them. Aim of this section is to study at a more quantitative level three specific  models, two with $w_0\neq 0$
and one with $w_0=0$. Most of the analysis here is performed numerically, because  the exponential nature of the superpotential and the smallness of the $D_X$ term at the minimum require a detailed knowledge of the location of the vacuum, in particular in the moduli directions. In order to appreciate this point, we will report in tables 1, 2 and 3 various quantities of interest computed starting both by the exact numerical vacuum
and the approximate analytical one. The latter is found along the lines of subsection 2.1. We start from the SUSY configuration
for $\phi$, $U$ and $Z$ given by eqs.(\ref{phi0}) and (\ref{UZsusy}) for $w_0\neq 0$,
and eqs.(\ref{phi0}), (\ref{Ususyw0}) and (\ref{Zsusyw0}) for $w_0=0$ . We then expand $V_0$, taking $V_{up-lift}$ as a perturbation, around the SUSY vacuum, keeping only the linear terms in $\Delta U$, $\Delta Z$ and $\Delta \phi$. In this way we get what we denoted by $U_0$, $Z_0$ and $\phi_0$. We finally compute the VEV of $\chi$ as explained below eq.(\ref{chiscal}).

Notice that even the numerical search of exact minima in presence of the $D_X$-term is not straightforward.
The $D_X$-term is naturally of order one when slightly off-shell, and thus much bigger than the typical
values of the $F$ terms, namely one has $V_D\gg V_F$ and all the energy of the system is dominated by $V_D$, hiding completely the stabilization of the moduli encoded in $V_F$.  On the contrary, in the heavy gauge field approximation, $V_D\ll V_F$  at the minimum, being $O(F^4)$, see eq.(\ref{Datvacuum}), which means that a severe fine-tuning takes place in $V_D$ at the
minimum.  We have been able to circumvent this problem using a linear combination of
the equations of motion for $\phi$ and $\chi$ to solve for $D_X$ in terms of $F$-terms and their derivatives, and replace the result
in these.  We also found useful, instead of solving the equation of motion for $\phi$, to impose that the expression found for $D_X$ to be equal to its expression (\ref{Dxterm}) in terms of the fields. The numerical vacua so obtained turns out to be stable and the resulting $D_X$ and $F$-terms always satisfies the consistency condition (\ref{FDrel}). 

\subsection{IIB Model}

The first model we consider is based on an orientifold compactification of type IIB string theory on a Calabi-Yau  3-fold  obtained
as an hyper-surface in  $\textbf{CP}^4$, namely $\textbf{CP}^4_{[1,1,1,6,9 ]}$. This Calabi-Yau has $h^{1,1} =2$ and $h^{2,1}=272$ K\"ahler and complex structure moduli, respectively. See \cite{Denef:2004dm} for details. In the spirit of \cite{KKLT}, we assume
here that a combination of NSNS and RR fluxes stabilize the dilaton and complex structure moduli supersymmetrically.
Once integrated out, these fields just give rise to a constant superpotential term $w_0$. 
We do not specify the detailed string construction which might give rise to the superpotential (\ref{fullWuv}).   We generally assume that $D_7$ branes (and $O_7$ planes) must be introduced to generate the non-perturbative superpotential terms in (\ref{fullWuv}), as well as the non-linear transformation under $U(1)_X$ of the modulus $U$  \cite{Haack:2006cy,Jockers:2005zy}.  
We will neglect in the following possible open string moduli and consider the dynamics of the two K\"ahler moduli only, identifying them with the two moduli $U$ and $Z$.
We have now to specify the explicit form of the various terms entering in the K\"ahler potential (\ref{fullKuv}).
The purely moduli-dependent function $K_M$ is known. In the usual approximation of neglecting flux effects, it 
takes the form \cite{Denef:2004dm}
\be
K_M=-2 \log {\rm Vol} ~,~~{\rm Vol}= \frac{1}{9\sqrt{2}}\left(\left(\frac{U+\bar U}{2}\right)^{3/2}-\left(\frac{Z+\bar Z}{2}\right)^{3/2}\right),
\label{IIBkaehler}
\ee
where ${\rm Vol}$ is the volume of the Calabi-Yau manifold.
We do not specify the modular functions $\alpha_{1,2}$ for the mesons $M_{1,2}$ since they do not play any role in the limit
where the mesons are supersymmetrically integrated out. The modular functions for $\phi$ and $\chi$, $\alpha_\phi$ and $\alpha_\chi$ in eq.(\ref{fullKuv}), are instead relevant but are generally difficult to derive and depend on the underlying string construction. We assume here the following ansatz:
\be
\alpha_\phi=\alpha_\chi = \frac{(Z+\bar Z)}{{\rm Vol}}\,,
\label{KahlerEx}
\ee
which is simple enough, but not totally trivial. It should be stressed that there is nothing peculiar in the (arbitrary) choice we made in eq.(\ref{KahlerEx}). Any other choice will be fine as well, provided that $\alpha_\chi$ is not too small. Indeed, according to eq.(\ref{chiscal}),
$\chi_0\sim \alpha_\chi^{-1/2}$ and a sufficiently small $\alpha_\chi$ can lead to a breakdown of our analysis based on an expansion in $\chi$. We have now to specify the various parameters entering in the hidden superpotential (\ref{fullWuv}) and the gauge kinetic functions
(\ref{gaugekinetic}).  Their choice is somehow arbitrary, but we require that $U$ and $Z$ and the volume of the Calabi-Yau to be sufficiently large to trust the classical SUGRA analysis. We take, for $\eta_1=-\eta_2=-1$,\footnote{What actually matters are the values of the phenomenological parameters (\ref{abcoeff}),
which do not uniquely fix the microscopical ones, as is evident from  eq.(\ref{abcoeff}).
The choice (\ref{IIBparameters}) is purely illustrative. The same comment also applies to the next
two examples below.}  
\bea
N_1= 40,~~N_{1f}=4,~~q_1=1,~c_1=1,~p_1=0,&&\!\! n_1=\frac{1}{4\pi}, ~ m_1=0,~\eta_{1,U}=\eta_{1,Z}=0,\cr
N_2= 25,~~N_{2f}=1,~~q_2=0,~c_2=1,~p_2=0,&&\!\!~ n_2=0,~ m_2=\frac{1}{4\pi},~\eta_{2,U}=\eta_{2,Z}=0,\cr
 w_0= 9\times10^{-14}, ~~Y= 4.2\times 10^{-5},~~q_\chi= 6, && \!\! \gamma_Z=\frac{1}{60},~\gamma_U=0, ~~n_X =\frac{1}{4\pi}\,.
 \label{IIBparameters}
\eea
\TABLE[ht]{

\begin{tabular}{cc}

\begin{tabular}{|l|| l | l | }
\hline
  & Numerical & Analytical \\
\hline
$\langle U\rangle$     & $232$                   & $230$  \\
$\langle Z\rangle$     & $148$                   & $146$   \\
$\langle \phi\rangle$  & $6.2\times 10^{-2}$     & $6.3\times 10^{-2}$ \\
$\langle \chi\rangle$  & $2.2\times 10^{-4}$     & $1.3\times 10^{-4}$ \\
$\langle F_U \rangle$                  & $-4.0\times 10^{-16}$   & $-2.3\times 10^{-16}$   \\
$\langle F_Z \rangle$                  & $4.2\times 10^{-16}$    & $2.3\times 10^{-16}$ \\
$\langle F_\phi \rangle$               & $8.5\times 10^{-15}$    & $5.1\times 10^{-15}$   \\
$\langle F_\chi \rangle$               & $2.1\times 10^{-13}$    & $2.2\times 10^{-13}$  \\
$\langle D_X \rangle$                  & $1.5\times 10^{-26}$   & $1.6\times 10^{-27}$ \\
$\langle V\rangle$     & $9.2 \times 10^{-33}$   & $2.2 \times 10^{-32}$\\
\hline
\end{tabular}
\begin{tabular}{|l|| l | l | }
\hline
 & Numerical & Analytical \\
\hline
$m_{3/2}$                     & $1.6$              & $1.6$         \\
$m_{\phi}$                    & $7.3\times 10^{13}$ & $7.4\times 10^{13}$\\
$m_{{\rm Re}(\chi)}$          & $242$               & $254$            \\
$m_{{\rm Im}(\chi)}$          & $241$               & $253$             \\
$m_{{\rm Im}(\tilde U)}$      & $140$               & $176$             \\
$m_{{\rm Re}(\tilde Z)}$      & $97$                & $156$           \\
$m_{{\rm Im}(\tilde Z)}$      & $95$                & $105$             \\
$m_{{\rm Re}(\tilde U)}$      & $61$                & $93$          \\
\hline
\end{tabular}
\end{tabular}
\label{tab.1}
\caption{VEVs, masses and scales for the IIB model with $w_0\neq 0$ and 
parameters given by eq.\eqref{IIBparameters}. Expectation values are expressed in (reduced) Planck units and
masses in TeV units. $\tilde U\sim U+Z$ and $\tilde Z\sim U-Z$ stand for the (approximate) eigenvector mass states.  
The definitions are slightly different in the numerical and analytical cases due to the diagonalization of the kinetic terms.}
}
The exponential moduli dependence in the up-lifting term $V_{up-lift}$ is supposed to arise from some non-perturbative effect, such as stringy instantons \cite{Blumenhagen:2006xt}.  The supersymmetric vacuum when the $\chi$-sector is turned off is 
\be
U_{SUSY}= 229~,~~ Z_{SUSY}= 145~,~~\phi_{SUSY}=6.3\times 10^{-2}\,.
\label{SUSYIIB}
\ee

We report in table \eqref{tab.1} (left panel) the location of the non-SUSY vacuum, as exactly found numerically and analytically by linearly expanding around the SUSY solution (\ref{SUSYIIB}), as well as the $F$-terms, $D_X$ and the potential $V$ at the minimum.
It can be seen that $U$, $Z$ and $\phi$ are well reproduced analytically, whereas $\chi$ is not, since its VEV is exponentially sensitive to the values of $U$ and $Z$ by means of the non-perturbative terms in $W_{eff}$. For the same reason $F_U$, $F_Z$ and $F_\phi$ are only roughly reproduced.  $F_\chi$, instead,  is better estimated since
it essentially depends on $Z$ only through the mild exponential appearing in the first term in eq.(\ref{yukawauv}). 
The $D_X$ term is also well reproduced because at leading order it is governed by $F_\chi$ only, see eq.(\ref{Datvacuum}).
In table \eqref{tab.1} (right panel) the gravitino and all the scalar masses are reported, in TeV units. For simplicity of presentation, we have not written the precise linear combination of mass eigenvectors, but just the main components in field space.
As anticipated, there is a hierarchy of scales. Fixing the overall scale such that $m_{3/2}\simeq O$(1)TeV, the field $\phi$ (and the $U(1)_X$ gauge boson $A_X$) is ultra-heavy, whereas the moduli and $\chi$ have masses  O($100$) TeV.  Like for the F-terms, the masses which do not directly depend on the strong dynamics, namely $m_{3/2}$, whose mass is governed by $w_0$, and $m_\phi$, whose mass is well approximated by eq.(\ref{MPHI}), are well predicted analytically. 
In agreement with our general observations, $1/(a_1 U) \simeq 1/(b_2 Z) \simeq 1/37\equiv \epsilon$. 
 It is easy to check that the values of $F_\phi$, $F_U$ and $F_Z$ reported in table \eqref{tab.1} agree with the scaling behaviors predicted by eq.(\ref{FtermsSca}). 
As observed in subsection 2.1, the stability of the system requires a low value for $\gamma_Z$ and hence the exponential dependence on $Z$ of the coupling $Y(U,Z,\phi)$ does not help much in  getting a not so small  $Y$. This constraints us to choose a rather large value of the $U(1)_X$ charge of $\chi$, $q_\chi=6$, although such choice might not naturally appear in simple $D$-brane constructions. 
We can also compute the universal gaugino masses at the high scale, assuming $f_v =f_2 = m_2 Z$. 
We get
\be
m_g \simeq 380 \; {\rm GeV}\,,
\ee
which is roughly one quarter the gravitino mass.
As explained before, we assume $U(1)_X$ neutral visible matter fields, so that the non-holomorphic soft scalar masses $m\sim m_{3/2} \sim O(1)$ TeV, instead of
$m\simeq g_X \sqrt{D_X q_{iv}} \simeq 70 \,{\rm TeV}\, \sqrt{q_{iv}}$, valid for $U(1)_X$ charged fields.

We have finally considered the reliability of the SUGRA approximation by considering the $\alpha^\prime$ correction
appearing in $K_M$. For type IIB orientifolds, this is known to be
\cite{Becker:2002nn}
\be
{\rm Vol}\rightarrow {\rm Vol}+\frac{\xi}{2g_s^{3/2}}, \ \ \ \ \ \ \xi =  -\frac{\chi(M) \zeta(3) }{2(2\pi)^3}\,,
\label{alphaprimeII}
\ee
with $\chi(M)$ the Euler characteristic of the Calabi-Yau and $\zeta(3)=\sum_{n=1}^\infty 1/n^3 \simeq 1.2$.
In writing eq.(\ref{alphaprimeII}), we have frozen the dilaton field $S$, which is taken non-dynamical and stabilized by fluxes to some value
$g_s={\rm Re}\,S$. We have studied how the correction (\ref{alphaprimeII}) roughly affects the model given by the input parameters (\ref{IIBparameters}), which are kept all fixed with the exception of $Y$, which is tuned to get an approximately Minkowski vacuum. A sizable correction is expected when $ \xi/2g_s^{3/2}\sim {\rm Vol}$. This is indeed the case, although
we have numerically checked that the $\alpha^\prime$ correction is non-negligible already when it is $O({\rm Vol}/10)$.
In our example $\chi(\textbf{CP}^4_{[1,1,1,6,9 ]})=-540$ and for the vacuum shown in  table \eqref{tab.1},
roughly speaking, $g_s\gtrsim 1/10$ in order to trust the SUGRA analysis and neglect the correction (\ref{alphaprimeII}).

\subsection{Heterotic Model}

The second model we consider is inspired by a  generic compactification of perturbative heterotic string theory on a Calabi-Yau 3-fold.
In such a case, we identify $U$ and $Z$ with the dilaton and the universal K\"ahler modulus, respectively.\footnote{In a more conventional notation the dilaton and universal K\"ahler modulus are denoted by $S$ and $T$, respectively. }
Contrary to the IIB case, unfortunately we dot not still have a scenario to stabilize in a controlled way the complex structure moduli
in the heterotic string. In addition to that, the presence of a small non vanishing constant superpotential term $w_0$ 
is known to be not easily produced. Indeed, a flux for $H$ induces a constant term in the superpotential \cite{Dine:1985rz},
but being the $H$ flux quantized, such a constant term is typically of order one in Planck units \cite{Rohm:1985jv}.
In the spirit of our bottom-up SUGRA approach, we do not look for a microscopic explanation for $w_0$. It might be the left-over term of the $F$ and $D$ terms vanishing conditions for all the extra fields which are supposed to occur in any realistic model, or else the left-over of a flux superpotential of an heterotic string compactified on generalized half-flat manifolds and non-standard embedding, which has been argued to admit quantized fluxes resulting in a small $w_0$ \cite{deCarlos:2005kh}.
The classical K\"ahler potential for the moduli is known to be \cite{Witten:1985xb}
\be
K_M=- \log(U+\bar U)-3 \log(Z+\bar Z)\,.
\ee
The modular functions $\alpha_\phi$  and $\alpha_\chi$ are now functions of $Z$ only and typically expected to be of the form 
 $(Z+\bar Z)^{-n}$, with $-1 \leq n \leq 0$.  We take here $n=-1/3$, so that
 \be
 \alpha_\phi = \alpha_\chi =\frac{1}{(Z+\bar Z)^{1/3}}\,.
 \ee
The parameters entering in the gauge kinetic functions (\ref{gaugekinetic}) and the superpotential (\ref{fullWuv})
 are quite more constrained with respect to the IIB case. At tree-level $f_i = f_X = U$. A possible $Z$-dependence can
 (and generally does) occur only at loop level by means of moduli-dependent threshold corrections. The exponential moduli dependence of the couplings $Y$ and $c_i$ is assumed to arise from world-sheet instantons and hence they depend on $Z$ only. Finally, the
size of the hidden gauge groups is bounded. We will look for vacua with  $U,Z\gtrsim 1$, which are 
on the edge of perturbativity, but lie in the phenomenologically most interesting region in moduli space
in perturbative heterotic theory. In fact, one should require  ${\rm Re}\, U \sim 2$ for a successful SUSY GUT model, but since
the 10-dimensional string coupling is given by $g_s=\sqrt{Z^3/U}$ \cite{Witten:1985xb}, perturbativity of the 10d
heterotic string also requires $Z^3< U$ and hence $Z \gtrsim 1$.
We then take 
\bea
N_1=5,~N_{1f}=4, && q_1=1,~~c_1=\frac 12,~~\eta_{1,Z}=2\pi+\frac 34,~\eta_{1,U}=0,~m_1=0,~n_1=1,~p_1=0\nn \\
 N_2=4,~N_{2f}=2, && q_2=2,~~c_2=1,~~~\eta_{2,Z}=0,~~~~~~~~\eta_{2,U}=0,~~m_2=0,~n_2=1,~p_2=0\nn \\
 w_0=3 \times 10^{-15}, ~ && q_\chi=10,~Y=3.1\times10^{-3},~~~~~~~~~~~~\gamma_Z=2\pi,~\gamma_U=0,~n_X=1\,.
 \label{HETPARA}
\eea
The supersymmetric vacuum is at
\be
U_{SUSY}= 1.76,~ Z_{SUSY}= 1.19,~\phi_{SUSY}=0.14\,.
\label{HETSUSY}
\ee
\TABLE[ht]{
\begin{tabular}{c c}

\begin{tabular}{|l|| l | l |}
\hline
 & Numerical & Analytical \\
\hline
$\langle U\rangle$     & $1.78$               & $1.78$  \\
$\langle Z\rangle$     & $1.20$               & $1.20$   \\
$\langle \phi\rangle$  & $0.14$               & $0.14$ \\
$\langle \chi\rangle$  & $-3.2\times 10^{-4}$  & $-2.0\times 10^{-4}$ \\
$\langle F_U\rangle$                  & $-1.9\times 10^{-16}$ & $-1.5\times 10^{-16}$   \\
$\langle F_Z\rangle$                  & $-1.4\times 10^{-15}$& $-1.0\times 10^{-15}$  \\
$\langle F_\phi\rangle$               & $-2.1\times 10^{-17}$ & $-7.2\times 10^{-18}$  \\
$\langle F_\chi\rangle$               & $4.0\times 10^{-15}$  & $4.2\times 10^{-15}$   \\
$\langle D_X\rangle$                  & $5.4\times 10^{-28}$ & $5.9\times 10^{-28}$  \\
$\langle V\rangle$     & $1.3\times 10^{-32}$  & $1.7 \times 10^{-32}$  \\
\hline
\end{tabular}

&

\begin{tabular}{|l|| l | l |}
\hline
 & Numerical  & Analytical \\
\hline
$m_{3/2}$                     & $1.0$             & $1.0$         \\
$m_{\phi}$                    & $3.1\times10^{14}$ & $3.1\times10^{14}$\\
$m_{\chi}$                    & $191$              & $200$            \\
$m_{{\rm Re}(U)}$             & $98$               & $115$             \\
$m_{{\rm Im}(U)}$             & $87$               & $107$             \\
$m_{{\rm Im}(Z)}$             & $56$               & $61$             \\
$m_{{\rm Re}(Z)}$             & $42$               & $52$             \\
\hline
\end{tabular}
\end{tabular}
\label{tab.2}

\caption{VEVs, masses and scales for the heterotic model with $w_0\neq 0$ and
parameters given by eq.\eqref{HETPARA}. Expectation values are expressed in (reduced) Planck units and
masses in TeV units.}
}
In the heterotic case, being the tree-level gauge kinetic functions $U$-dependent only ($m_{1,2}=0$), 
one cannot have both  $\eta_{1,Z}=\eta_{2,Z}=0$, since this would lead to the vanishing of the effective parameters
$b_{1,2}$ defined in eq.(\ref{abcoeff}), which is unacceptable. The particular choice of $\eta_{1,Z}$ in eq.(\ref{HETPARA}) is
required to fix $Z$ in the small window $1\lesssim Z\lesssim U$, but of course there are several ways to achieve it in terms of the microscopic parameters,  given that  what matters 
are the effective ones defined in eq.(\ref{abcoeff}).  The asymmetry between $U$ and $Z$ in this heterotic inspired model
does not allow to straightforwardly use the general scalings discussed in section 2.1.
We do not report the analytical more elaborated analysis which is now required.
One can nevertheless check that the general scalings (\ref{deltaX}) and (\ref{FtermsSca})  still 
give the rough order of magnitude estimate for the shifts of the fields and the $F$ terms by taking $\epsilon \simeq 1/(a_1 U) \simeq 1/28$.  
We report in table \eqref{tab.2} (left and right panel) the location of the non-SUSY vacuum, the
$F$-terms, $D_X$, the potential and the masses for the scalars and the gravitino of the model.
As in the IIB model before, $\chi$ is not analytically reproduced to a good accuracy, being exponentially
sensitive to the values of $U$ and $Z$. Similarly for $F_U$, $F_Z$ and $F_\phi$. 
The hierarchy of scales appearing in the gravitino and  scalar masses are the same as in the Type IIB model.
The large $U(1)_X$  charge of $\chi$, $q_\chi = 10$,  allows for a natural explanation of the smallness of the
up-lifting term without using tiny values for $Y$, as was done in (\ref{IIBparameters}) in the previous example.
The tree-level universality of the heterotic holomorphic gauge kinetic functions (for level one Kac-Moody gauge groups, assumed here)
fixes $f_v=U$ and hence the universal high scale gaugino masses are calculable and read 
\be
m_g \simeq 240 \; {\rm GeV}\,.
\ee
Unfortunately it is not possible now to assume all visible matter fields to be $U(1)_X$ neutral, since $f_v=U$.
For the $U(1)_X$ charged fields we get now $m\simeq g_X \sqrt{D_X q_{iv}} \simeq 42 \,{\rm TeV}\, \sqrt{q_{iv}}$. 

The vacuum reported in table (\ref{tab.2}) is barely perturbative since the associated 10d string coupling constant $g_s \simeq 1$.
One can explicitly see that the situation is similar in the $\alpha^\prime$ expansion, due to the low values of the moduli,
by considering again the universal $\alpha^\prime$ correction to the K\"ahler potential for $Z$,
which now reads \cite{Candelas:1990rm}
\be
(Z+\bar Z)^3\rightarrow (Z+\bar Z)^3+4\xi,
\label{alphaprimeHet}
\ee
with $\xi$ defined as in eq.(\ref{alphaprimeII}).  We find that the $\alpha^\prime$ correction (\ref{alphaprimeHet}) is generally negligible for $|\xi|\lesssim O(1/10)$
and are deadly for $|\xi|\gtrsim O(1)$, with the impossibility of achieving a Minkowski vacuum ($\xi<0$) 
or the appearance of tachyons ($\xi>0$). In the range $O(1/10)\lesssim |\xi| \lesssim O(1)$ the qualitative properties of the model are unaffected, but the numerical values reported in table (\ref{tab.2}) get corrections of order 100\%. Considering that for $|\chi(M)|\sim 10^2$, a typical value for Calabi-Yau manifolds, $|\xi|\sim O(1)$, it is clear that the phenomenologically interesting region $U,Z\gtrsim 1$ is barely calculable,
as we anticipated.
 
\subsection{IIB Model with $w_0=0$}

As we have seen, the requirement of a natural $Y$ favors the choice $\gamma_U\neq 0$ for models with $w_0=0$. In an heterotic context, where $U$ is the dilaton, we would be forced to invoke exotic non-perturbative couplings. In addition to that,  hierarchies would also appear in the mesonic Yukawa couplings $c_i$. The upper bound on the gauge groups leads to a lower bound on $a$, $a\gtrsim 15$ and $a_1-a_2\sim 3$. Using eq.(\ref{Ususyw0}), it is easy to see that  the phenomenological requirement $U\sim 2$ constraints $A_1/A_2\sim 10^3$. In light of eq.\eqref{abcoeff}, this hierarchy in $A_1/A_2$ typically induces an even larger hierarchy in the
microscopical Yukawa couplings $c_i$, unless  one assumes a hierarchy between them due to, say, world-sheet instantons.
For these reasons, as a specific example of model with $w_0=0$, we again opt here for a
type IIB model on $\textbf{CP}^4_{[1,1,1,6,9 ]}$, with $U$ and $Z$ the two K\"ahler moduli of 
the Calabi-Yau manifold. No bound on the gauge group  arises 
and the situation seems more favorable. For simplicity we take now trivial K\"ahler potentials for $\phi$ and $\chi$, namely
\be
\alpha_\chi  =  \alpha_\phi= 1\,.
 \ee
The input parameters are as follows:
\bea
N_1= 30,~~N_{1f}=2,~~q_1=2,~c_1=1,~~p_1=\frac{18}{50},&&\!\!~ n_1=\frac{1}{4\pi}, ~ m_1=0,~~~\eta_{1,U}=\eta_{1,Z}=0,\cr
N_2= 29,~~N_{2f}=2,~~q_2=2,~c_2=2,~~p_2=0,\,\,~&&\!\!~ n_2=\frac{1}{4\pi},~ m_2=0,~~~\eta_{2,U}=\eta_{2,Z}=0,\cr
N_3= 11,~~N_{3f}=1,~~q_3=0,~c_2=1,~~p_3=0,\,\,~&&\!\!~ n_3=0,~~~~m_3=\frac{1}{4\pi},~\eta_{3,U}=\eta_{3,Z}=0,\cr
Y=\frac{867}{5000}~,~~~~~~~q_\chi= 2,~n_X =\frac{1}{4\pi},&& \!\! ~\gamma_Z=0,~~~\gamma_U=\frac{1}{6}\,. 
 \label{RT3parameters}
\eea
The supersymmetric vacuum is at
\be 
U_{SUSY}= 136,  \ \ Z_{SUSY}= 63,  \ \ \phi_{SUSY}=0.10,
\ee 
and the exact non-SUSY vacuum, and its properties, is reported in table\eqref{tableRT}.
Notice that the model is quite constrained. Given $N_{1,2}$, $N_{1f,2f}$ and $q_{1,2}$,  for mesonic Yukawa couplings $c_i\sim O(1)$, the gauge kinetic functions are essentially fixed by  the requirement of low-energy SUSY. A constant threshold correction, appearing in (\ref{RT3parameters}), can be avoided by allowing a mild tuning between $c_1$ and $c_2$. The value of the coupling $Y$ is fixed by the flat condition.  As expected from the general arguments of section 4, the gauginos are now light and indeed we
get, for $f_v = f_3 = Z/(4\pi)$, 
\be
m_g \simeq 270 \; {\rm GeV}\,,
\ee
which is less than one order of magnitude smaller than $m_{3/2}$. We can take $U(1)_X$ neutral visible matter fields, so that the non-holomorphic soft scalar masses 
$m\simeq m_{3/2}$, instead of $m\simeq g_X \sqrt{D_X q_{iv}} \simeq 80 \,{\rm TeV}\, \sqrt{q_{iv}}$, valid for $U(1)_X$ charged fields.
\TABLE[ht]{
\begin{tabular}{cc}

\begin{tabular}{|l|| l | l | }
\hline
 & Numerical & Analytical \\
\hline
$\langle U\rangle$     & $136.0$                  & $136.0$  \\
$\langle Z\rangle$     & $63.3$                   & $63.3$   \\
$\langle \phi\rangle$  & $0.101$     & $0.101$ \\
$\langle \chi\rangle$  & $-3.6\times 10^{-6}$     & $-7.6\times 10^{-7}$ \\
$\langle F_U \rangle$                  & $-2.7\times 10^{-17}$   & $-2.6\times 10^{-17}$   \\
$\langle F_Z \rangle$                  & $-7.2\times 10^{-18}$    & $-8.1\times 10^{-18}$ \\
$\langle F_\phi \rangle$               & $1.5\times 10^{-16}$   & $1.6\times 10^{-16}$   \\
$\langle F_\chi \rangle$               & $2.0\times 10^{-13}$    & $2.0\times 10^{-13}$  \\
$\langle D_X \rangle$                  & $1.2\times 10^{-26}$    & $1.2\times 10^{-27}$ \\
$\langle V\rangle$     & $2.4 \times 10^{-32}$   & $2.4\times 10^{-32}$\\
\hline
\end{tabular}
\begin{tabular}{|l|| l | l | }
\hline
 & Numerical & Analytical  \\
\hline
$m_{3/2}$                     & $3.31$              & $3.31$         \\
$m_{\phi}$                    & $1.1\times 10^{14}$ & $1.1\times 10^{14}$\\
$m_{{\rm Im}(\tilde U)}$      & $6.3\times 10^3$    &$6.3\times 10^3$            \\
$m_{{\rm Re}(\tilde U)}$      & $6.3\times 10^3$    &$6.3\times 10^3$          \\
$m_{{\rm Re}(\tilde Z)}$      & $229$               & $229$             \\
$m_{{\rm Im}(\tilde Z)}$      & $212$               & $212$           \\
$m_{{\rm Im}(\chi)}$          & $160$               & $160$             \\
$m_{{\rm Re}(\chi)}$          & $160$               & $160$          \\
\hline
\end{tabular}
\end{tabular}
\label{tableRT}
\caption{VEVs, masses and scales for the IIB model with $w_0= 0$ and
parameters given by eq.\eqref{RT3parameters}. Expectation values are expressed in (reduced) Planck units and
masses in TeV units.  $\tilde U\sim U+Z$ and $\tilde Z\sim U-Z$ stand for the (approximate) eigenvector mass states.}}

The analytical and numerical values in this case agree with very good accuracy thanks to the smallness of $\chi$. We have $1/(b_3 Z)\simeq 1/36\sim 1/(a_{1,2} U)\simeq 1/30$
and the expected scalings for the shifts of the fields and the $F$ terms are satisfied.
The value of $\chi$ as would be roughly predicted by eq.(\ref{chiw0}) is about one order of magnitude
bigger than its actual value, due to  accidental numerical factors such that $q_\chi \chi F_\chi\sim \phi F_\phi/10$.
Notice how this hybrid model, where the KKLT-like and racetrack stabilizations work together, has all the appealing features we were looking for.  The stabilization via the racetrack mechanism of $U$ allows  to get a natural Yukawa coupling $Y$, and the stabilization of $Z$ after the uplifting generates acceptable  gaugino masses and avoid the problem of having too heavy scalar soft masses.

We have numerically checked the reliability of the classical SUGRA analysis by looking at the
$\alpha^\prime$ correction (\ref{alphaprimeII}). As expected, the correction becomes sizable for $|\xi|/g_s^{3/2}\gtrsim {\rm Vol}$, i.e, $g_s\sim 1/25$ and
is essentially negligible for $|\xi|/g_s^{3/2}\lesssim {\rm Vol}/10$. This results on a mild lower bound on the string coupling: $g_s\gtrsim 1/6$.  Contrary to the case with $w_0\neq0$ where the possibility of getting a Minkowski vacuum is lost for $g_s$ smaller than the bound, no dramatic consequences appear now, in the sense that the corrections are only quantitative, but the non-SUSY Minkowski vacuum  
is still there, even  for $g_s\sim 1/30$.\footnote{In this and in the previous IIB example with $w_0\neq0$,
the assumed decoupling between K\"ahler and dilaton moduli stabilization will give rise to a further constraint on $g_s$, since the $\alpha^\prime$ correction (\ref{alphaprimeII})  introduce a mixing between the two sectors.}

\section{Conclusions}\label{concl}

We have shown in this paper how to get realistic, string inspired, SUGRA Minkowski/dS vacua with two moduli, where the SUSY breaking
mechanism is induced by a FI term appearing in the $D$ term of a would-be anomalous $U(1)_X$ gauge symmetry. 
Like in the original Fayet model, the FI coupling and the presence of a superpotential term $\chi \phi^{q_\chi}$ for
two charged fields $\phi$ and $\chi$  force the system to break SUSY. 
Being the FI term moduli-dependent,  the stabilization of the
moduli is a crucial (and anyway important) ingredient in the framework. The mechanism 
can be made stable and relatively natural by invoking a ratio of $U(1)_X$ charges between $\phi$ and $\chi$ by almost one order of magnitude.
When the moduli are stabilized without a constant superpotential term $w_0$, 
the mechanism is more robust and no bound on the $U(1)_X$ charges arises. The moduli masses are proportional to the scale of SUSY
breaking and hence a gravity mediation of SUSY breaking, with a gravitino mass of $O({\rm TeV})$, is preferred for cosmological reasons.

Studying the dynamics of two moduli, instead of one only, is not just an academic complication, because in this set-up
one modulus necessarily transforms under the $U(1)_X$ gauge symmetry, whereas the second can be taken neutral.
The two-moduli system is then the simplest scenario that can describe more realistic situations.
Moreover, we have shown how the presence of the second neutral modulus can considerably help in getting sufficiently heavy gaugino masses without getting at the same time very heavy scalar masses, a property which is not so common
in string compactifications.  

All our analysis has been based on the search of non-SUSY solutions starting from SUSY ones. 
It should be stressed that this does not imply that our vacua are small deformations of SUSY ones,
 since the small parameter $\epsilon$ is fixed by the Planck/weak scale hierarchy and cannot be continuously taken to be zero. 
On the other hand, we cannot exclude the existence of other interesting non-SUSY vacua which are not detectable with our approach.
We think that the above SUSY breaking mechanism, together with the stabilization of moduli by non-perturbative gauge dynamics,
is interesting, promising and can be made quite natural,  particularly when $w_0=0$. The next crucial step, as always when considering bottom-up SUGRA phenomenological models, would be to embed this mechanism in a full string theory set-up.

\section*{Acknowledgments}

This work is partially supported by the European Community's Human
Potential Programme under contracts MRTN-CT-2004-005104, by the Italian MIUR under contract 
PRIN-2005023102 and by INFN.

\end{document}